\documentclass[namedreferences]{solarphysics}
\usepackage[hyperref,optionalrh,solaromanenum,showbiblabels]{spr-sola-addons} 
\usepackage{graphicx}                    
\usepackage{natbib}
\usepackage{color}                       
\usepackage{breakurl}                         

\chardef\us=`\_

\begin{document}

\begin{article}

\begin{opening}

\title{Examining the Magnetic Field Strength and the Horizontal and Vertical Motions in an Emerging Active Region}

%
\author[addressref={aff1},corref,email={chlin@jupiter.ss.ncu.edu.tw}]{\inits{C.-H.}\fnm{Chia-Hsien}~\lnm{Lin}}
\author[addressref={aff2},email={commanderchen@gmail.com}]{\inits{Y.-C.}\fnm{Yu-Che}~\lnm{Chen}}
\address[id=aff1]{Graduate Institute of Space Science,
National Central University,
Chung-Li, Taoyuan, Taiwan}
\address[id=aff2]{Department of Atmospheric Science,
National Central University,
Chung-Li, Taoyuan, Taiwan}

%
\runningauthor{Lin and Chen}
\runningtitle{Examining an Emerging Active Region}


\begin{abstract}
Earlier observational studies have used the time evolution of 
emerging magnetic flux regions at the photosphere to infer their subsurface
structures, assuming that the flux structure does not change significantly
over the near-surface layer.
In this study, we test the validity of this assumption by comparing
the horizontal and vertical motions of an emerging active region.
The two motions would be correlated if the emerging structure is rigid.
The selected active region (AR) NOAA 11645
is not embedded in detectable preexisting magnetic field.
The observed horizontal motion is 
quantified by the separation of the two AR polarities
and the extension of the region.
The vertical motion is derived from the magnetic buoyancy theory.
Our results show that the separation of the polarities 
is fastest at the beginning with a velocity of  $\approx$~4~Mm hr$^{-1}$
and decreases to $\le$~1~Mm hr$^{-1}$ after the main growing phase
of  flux emergence.
The derived thick flux-tube buoyant velocity is between 1 and 3~Mm hr$^{-1}$ while
the thin flux-tube approximation results in an unreasonably high buoyant velocity,
consistent with the expectation
that the approximation is inappropriate at the surface layer.
The observed horizontal motion is not found to
directly correlate with
either the magnetic field strength or the derived buoyant velocities.
However, 
the percentage of the horizontally oriented fields
and
the temporal derivatives of the field strength and the buoyant velocity
show some positive correlations with the separation velocity.
The results of this study imply that
the assumption that the emerging active region
is the cross section of a rising flux tube
whose structure can be considered rigid as it rises through
the near-surface layer should be taken with caution.
\end{abstract}

%
\end{opening}

%
\section{Introduction} \label{sec:intro}
Sunspots and solar active regions are zones with
very concentrated magnetic fields
on the solar surface.
How these regions are formed is one of the fundamental problems
in the study of solar magnetism.
The formation process can be divided into two problems:
the first one is how the magnetic fields are transported to the surface,
and the second one is how the sunspots and active regions are formed from
these fields after they rise to the surface.

The first problem has been studied for more than half a century.
\citet{parker1955ApJ121} was the first to propose magnetic buoyancy
as a viable mechanism
to bring a strand of toroidal field to the surface.
Since then,
many works have been conducted to include more realistic effects
to the model of magnetic flux transport through the solar convective zone.
For instance,
\citet{Schuessler1979AA} 
considered the effects of differential rotation, flux loss
and convective motions,
and also derived a mathematical expression  for flux tubes with arbitrary sizes.
They found from their numerical simulation that the shape of the flux tube
changes during the rising.
\citet{Caligari_etal1995ApJ} included the effects of
spherical geometry and differential rotation
in their numerical simulations,
and reported that their results were consistent
with observed asymmetries, AR tilt angles and emergence latitudes.
\citet{Fan2008ApJ} examined the effects of magnetic twist and Coriolis force.
\citet{Weber_etal2011ApJ} applied a thin flux-tube model in a 
rotating spherical shell of turbulent convective flows.
The validation of these models has usually been based on
whether a model can qualitatively produce the
observed general properties such as
Hale's law, hemispheric tilt and active latitudes.
The detailed flux transport processes
implemented in these models, however, is much harder to verify
because the process takes place in the invisible solar interior.
The magnetic fluxes only become observable
after they emerge from the photosphere.
Some earlier studies used the temporal evolution of
emerging flux regions (EFRs)
or emerging active regions (EARs) as a means to probe
the invisible part of the flux-tube structure and dynamics.
Specifically,
they considered the time sequence of the observed EFRs and EARs
as snapshots of different layers of the rising magnetic structure from top down,
and developed methods to reconstruct the subsurface structures
\citep{Tanaka1991SoPh,Leka_etal1996ApJ,CZ2013ApJ}.
These studies were based on the assumption that the structure
does not change significantly as it crosses the surface,
which, however, has not been verified observationally,
and contradicts some recent simulation results
\citep[{\it e.g.},][]{RempelCheung2014ApJ}.

The second problem is
how active regions are formed
from the fields being brought to the surface.
Are they the result of a rising field structure intersecting the surface,
or
are they formed by turbulence and/or convergence of
previously emerged small fields \citep[{\it e.g.},][]{lsmp1998AA}?
Although in this case the process is observable,
it is difficult to decipher the actual physical mechanisms.
This is because the gas density and pressure change radically over
a very thin layer across the surface,
specifically, the plasma $\beta \equiv P_{\rm gas}/P_{\rm mag}$ 
changes from $\beta \gg 1$ in the convection zone to $\beta \approx 1$,
which means that the plasma and magnetic field interact dramatically
in this layer.
As a result, the fields are distorted, deformed, and 
restrengthened by the plasma motions at this near-surface layer.
In addition,
since the physical scales of many complex processes,
such as turbulence, flows, convection, and others, 
become comparable with the scale
of the magnetic flux tube,
simplifications such as the thin flux-tube approximation 
and the anelastic assumption
are no longer valid in this layer.
Recent simulations have shown that
the structure of the flux tube can be significantly changed
during flux emergence across the near-surface layer
\citep[{\it e.g.},][]{Rempel2011ApJ,RempelCheung2014ApJ}.
\citet{Brandenburg2011ApJ} numerically demonstrated that
a negative effective magnetic pressure instability (NEMPI)
can lead to the formation of bipolar regions.
A comprehensive review can be found in \citet{CI2014LRSP}.
These models, simulations, and assumptions should be verified before
being applied to infer the physics associated with EFRs and EARs.

An early study to test the magnetic buoyancy theory was conducted by
\citet{CW1987SoPh}.
They derived the buoyant velocities of 24 emerging bipoles based on
the magnetic buoyant force of the thick flux-tube approximation \citep{Schuessler1979AA},
and compared them with observationally determined separation velocities of
the bipoles.
They found no correlations between the two velocities.
However,
their study utilized line-of-sight (LOS) magnetograms,
and was based on the simplification
that the separation speed and buoyant speed of a bipole
can be considered constant during the emergence.
More recently, improved observational and analysis techniques have enabled
more accurate examination of the magnetic and velocity fields of EFRs and EARs.
\citet{lsmp1998AA} conducted a high spatial resolution examination of
the vector magnetic field and Doppler velocity field patterns
of three young bipolar active regions
in which flux emergence could still be detected.
They found that the emerging horizontal magnetic elements had field strength
from $\approx$~200 to 600~G and upward rising speeds $\approx$~1~km s$^{-1}$.
These horizontal field elements
rapidly moved away from the site of emergence to
become more vertically oriented,
after which their field strength increased to kilogauss values.
Based on their results,
they suggested that the EFRs were caused by the emergence of 
subsurface structures
rather than by organized flows near the surface.
\citet{ksl2003ApJ} examined the internal structure
of an emerging flux region.
They divided the EFR into three regions: the main bipolar region,
two small emerging bipoles near the well-developed leading sunspot of 
the main bipole, and the remaining part of the main bipolar region.
By comparing the evolution of the field strengths and filling factors of
the main bipole and the small emerging bipoles,
they deduced that the reorganization of the magnetic fields by 
convective collapse \citep{parker1978ApJ} and 
flux concentration is a possible mechanism
for the evolution of the EFRs.
The subject of these two studies was flux emergence in
preexisting active regions.
\citet{centeno2012ApJ} studied the emergence of two active regions
that were isolated and not embedded in preexisting fields.
They analyzed the relationship between the Doppler velocities and
the inclination angle, relative to LOS, of the magnetic field vectors
of all magnetic flux elements
in the emerging active regions.
They found that the fields connecting two polarities were horizontally oriented
with strong upward plasma velocities while
the magnetic fields in the footpoints were vertically oriented 
with downward plasma velocities.
In these studies, 
every magnetic flux element was treated as independent,
and their statistical analysis was the basis
to deduce the possible formation mechanism(s) of their selected regions.

The aim of the present article is
to test the assumption that an emerging active region
is formed by the emergence of a subsurface flux structure
whose dynamics is mainly governed by the magnetic buoyancy mechanism
and whose structure can be considered rigid as it moves across the 
near-surface layer.
For an ideal rigid structure,
its vertical and horizontal motions,
being the two perpendicular components of the total motion,
are expected to be correlated.
Our strategy is to choose an EAR that
qualitatively resembles the emergence of a single flux tube,
and check whether any dependence exists between 
its horizontal and vertical motions.
The vertical motion is represented by the buoyant velocity
as derived by \citet{CW1987SoPh}.
We calculate the buoyant velocities in both thin flux-tube and 
thick flux-tube approximations
to check whether the approximations are valid in the EAR.
The horizontal motion is measured by the changing rates of
the separation of the two polarities,
the ``extension'' of the region,
and the area of the entire region.
We define ``extension'' as the size of the region in 
the direction perpendicular to the line connecting the two polarities.

\section{Data} \label{sec:data}
The vector magnetograms used in this work are derived from
the data taken by the {\it Helioseismic and Magnetic Imager} instruments 
\citep[HMI:][]{HMI_2012SoPh275}
on board the {\it Solar Dynamics Observatory}
\citep[SDO:][]{SDO_2012SoPh275}.
The computation of the HMI vector field products
has been explained in detail by \citet{HMI_ME2014SoPh289},
and a brief description is given here for completeness.
The data are
full-disk Stokes polarization parameters of
the spectral line {\tt Fe I}~6173~\AA \ sampled at six equally spaced wavelengths.
The Stokes vectors are then processed by  
the Very Fast Inversion of the Stokes Vector (VFISV) code,
which is implemented using the
Milne-Eddington approximation model of the solar atmosphere.
Although the original observation cadence is 45~s,
the data are averaged over a 16-minute tapered window
during the VFISV computation,
rendering a cadence of 12~min.
The results will be referred to as ``ME'' hereinafter.
The ME vector magnetograms are 
full-disk images of the total field, $|B|$, inclination angle, $\theta_B$,
and azimuth angle, $\phi$.
The inclination angle is measured relative to the line-of-sight (LOS) direction,
with $0^{\circ}$ pointing out of the image and $180^{\circ}$ into the image. 
The azimuth angle is measured counterclockwise (CCW) from the solar North, 
and contains the 180$^\circ$ ambiguity.
In addition,
due to a combination of satellite and instrument roll angles, 
the ME data are rotated by approximately $180^\circ$ CCW relative
to the actual solar images  \citep{Sun2013arXiv}.
Since the analysis of this study is indifferent to such rotation,
the ME data were not corrected for the rotation,
except for visualization purpose.

In addition to the full-disk magnetograms,
HMI pipelines also provide active region patches (HARPs),
which are patches of emerging active regions
automatically detected and tracked. 
The detection and tracking
are based on photospheric LOS magnetograms and intensity images.
The $180^\circ$ azimuth ambiguity is resolved using
Metcalf's minimum energy method \citep{Metcalf1994SoPh,HMI_ME2014SoPh289}.
The geometric information from HARPs is subsequently used to calculate
various space-weather related quantities at each time step to
create Spaceweather HMI Active Region Patches (SHARPs)
\footnote{{\tt http://jsoc.stanford.edu/doc/data/hmi/sharp/sharp.htm}}.
Two types of vector magnetic field data are offered by SHARP:
data in the CCD coordinates ({\tt hmi.sharp\_720s}) and
data that are remapped using
a Lambert Cylindrical Equal-Area projection procedure
({\tt hmi.sharp\_cea\_720s}) 
such that the field quantities 
appear as if they were observed directly overhead
\citep{CalabrettaGreisen2002AA,Sun2013arXiv}.
The former data product, which will be referred to as ``SHARP'',
contains total field, inclination angle
and disambiguated azimuth angle.
The inclination angle is relative to the LOS,
and the azimuth angle is measured CCW from the solar North.
The mapped data series, which will be called ``CEA'' in this paper,
contains three field vectors,
$B_r$, $B_\theta$ and $B_\phi$,
where $\hat{r}$, $\hat{\theta}$, and $\hat{\phi}$ are
the basis vectors of a heliocentric spherical coordinate system
\citep{Sun2013arXiv}.

\section{Observations} \label{sec:obs}
\begin{figure} 
\centerline{\includegraphics[width=1.0\textwidth,clip=]{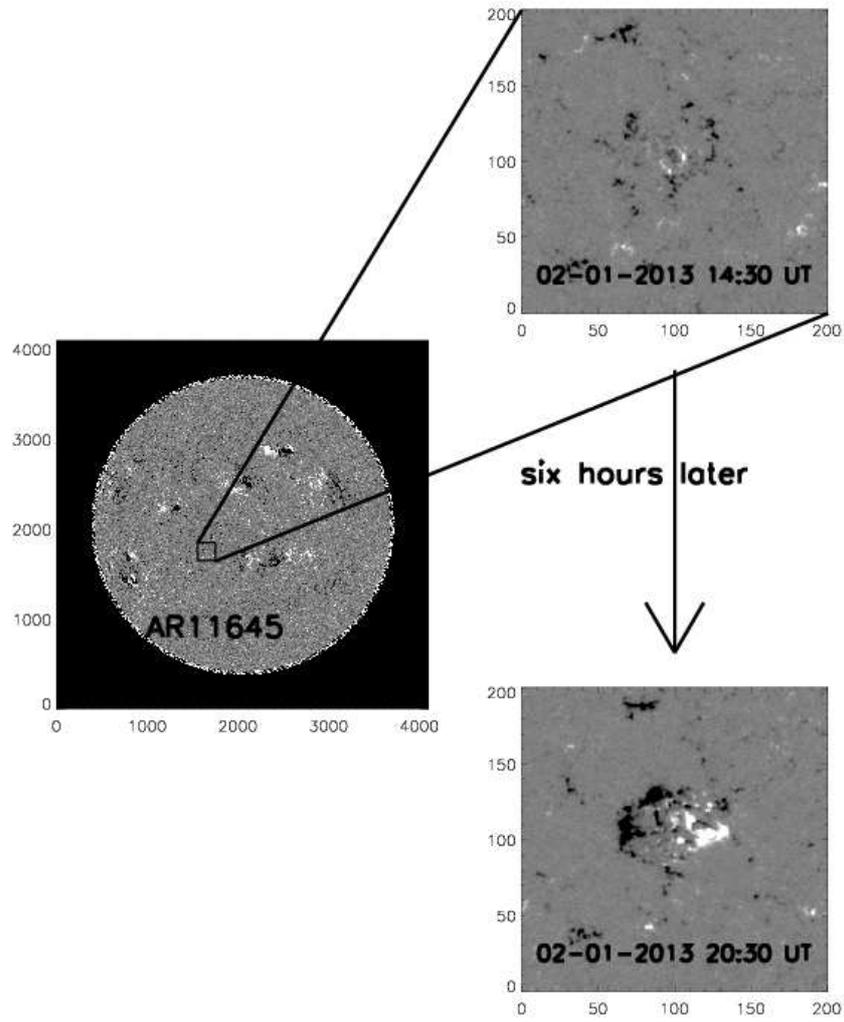}}
\caption{Emergence of AR11645 observed by SDO HMI line-of-sight magnetogram}
\label{fig:hmi}
\end{figure}

\begin{figure} 
\centerline{\includegraphics[width=1.0\textwidth,clip=]{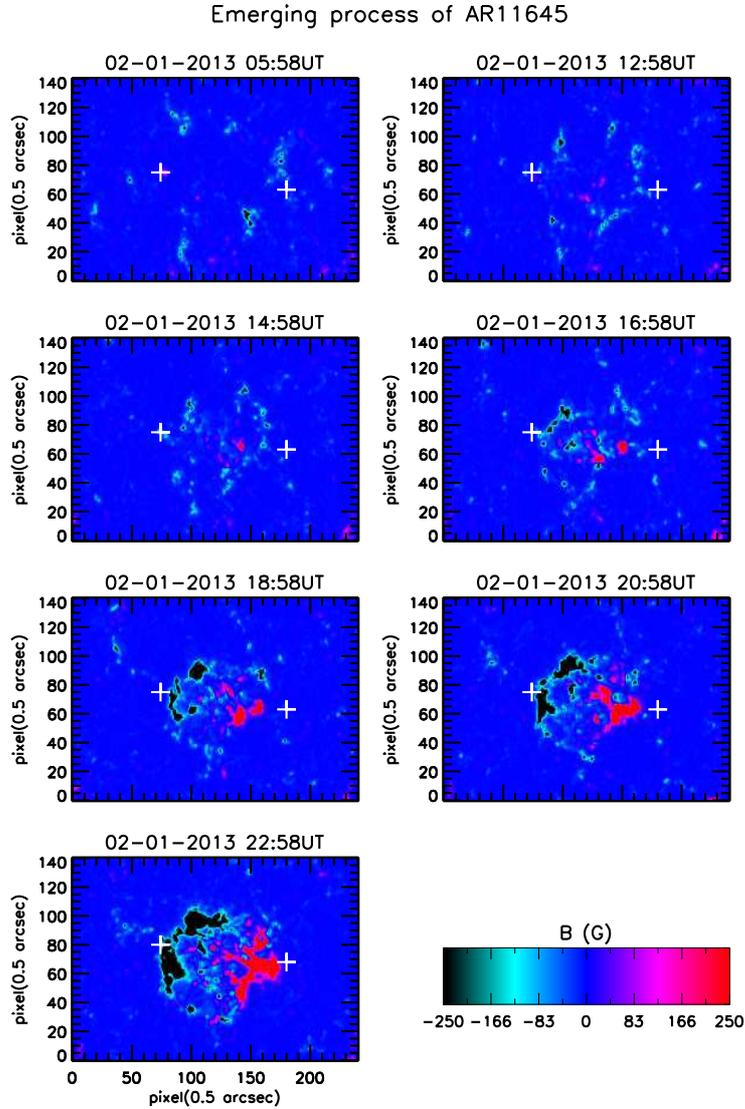}}
\caption{
Selected HMI magnetogram cuts to show the emergence 
of AR11645.
The observation times are indicated above the corresponding images.
The size of a pixel in both X and Y directions is 0.05 arcsec.
The two white crosses mark the approximate locations of the centers of 
the two polarities when the average total-field strength of 
the whole emerging active region reaches its maximum.
The plotted quantity is the line-of-sight component of magnetic field,
instead of the total field,
such that the two polarities can be distinguished.
}
\label{fig:emergence}
\end{figure}

The subject of this study is the emergence of
AR 11645.
Figure~\ref{fig:hmi} illustrates the location of its emergence
on the solar disk
as observed by HMI line-of-sight magnetogram.
The area of the flux-emergence region and the sign of the
magnetic flux concentration can be seen in more detail in
Figure~\ref{fig:emergence}. 
The two white crosses mark the approximate centers of the two main polarities
when the average total-field strength of 
the whole emerging active region reaches maximum.
The figure shows that
the area was initially free of preexisting magnetic field
and surrounded by a ring of predominantly negative field.
A small positive flux began to appear
between the two white crosses at approximately 12:58~UT on 2 January 2013.
As this positive region gradually increased its field strength,
multiple regions with mixed polarities rapidly appeared
between the two white crosses to fill the region.
Although many of these new fluxes did not
show a preferred polarity orientation when they first emerged,
they gradually oriented themselves along the East-West direction,
with the positive pole moving toward the right (solar West) and
the negative pole to the left,
leading the whole region to a bipolar configuration.
This is consistent with Hale's law.
The spatial range of the emergence analyzed in this study is
within $\pm 20^\circ$ in longitude,
from the perspective of the SDO instruments.

Because of the detection methodology,
the HARP automated procedure
begins to record the data of an emerging flux region only after
the LOS magnetic field of the region has become sufficiently strong
and a sunspot is seen in the intensity image.
Therefore, while the SHARP and CEA magnetograms
can provide the radial and horizontal components of the field without ambiguity,
they often do not include
the earliest stage of flux emergence 
when the field is mostly horizontal
and no sunspot has formed.
Since the selected emerging active region is located close to the disk center,
we expect the difference between the LOS and radial directions to be small.
Therefore, in this study, we mainly used ME data,
and incorporated CEA data to check for the errors caused by the
projection effect.

\section{Analysis} \label{sec:analysis}
\subsection{Coalignment of ME and CEA images}
The ME data are recorded in CCD coordinates with
a spatial size of 0.5 arcsec {\it per} pixel while
the CEA data are mapped to a cylindrical equal-area coordinate system
with a spatial size of 0.03 degree {\it per} pixel.
The standard CEA coordinates $(x,y)$ are related to the heliographic longitude
and latitude $(\phi,\lambda)$ as follows
\citep{CalabrettaGreisen2002AA,Sun2013arXiv}:
\begin{eqnarray}
x&=&\phi \nonumber \\
y&=& (180^{\circ}/\pi)\sin\lambda,
\label{eqn:cea0}
\end{eqnarray}
in which the reference point is the disk center.

To correct for the foreshortening effect in the CEA patches,
a spherical coordinate rotation is applied to Equation~(\ref{eqn:cea0})
to rotate the reference point to the patch center,
such that the result is an image observed directly from above
\citep{Sun2013arXiv}:
\begin{eqnarray}
x &=& {\rm arg}\left[\sin\lambda \sin\lambda_c + 
\cos\lambda \cos\lambda_c\cos(\phi-\phi_c), \; 
\cos\lambda\sin(\phi-\phi_c)\right] \nonumber \\
y &=& \sin\lambda\cos\lambda_c - \cos\lambda\sin\lambda_c\cos(\phi-\phi_c),
\label{eqn:cea_hmi}
\end{eqnarray}
where the function arg$[a,b]$ means $|\tan x|=|b/a|$,
where $x$ is in the same quadrant with the point $(a,b)$. 
$(x,y)$ in Equation~(\ref{eqn:cea_hmi}) is the CEA coordinate
of a point relative to the patch center, 
and is in radians.
$(\phi,\lambda)$ and $(\phi_c,\lambda_c)$ 
are the heliographic longitude and latitude of this point and the patch center,
respectively.

To ensure that the same region is cut from the ME and CEA data,
the pixel coordinates of the boundary 
({\it i.e.,} lower-left and upper-right corners)
of each ME image are first transformed into $(\phi,\lambda)$,
then transformed into the CEA coordinates using 
Equation~(\ref{eqn:cea_hmi}), 
and finally converted into the pixel locations:
\begin{eqnarray}
{\tt pix} &=& (x,y) \frac{180^{\circ}}{\pi}\frac{1}{0.03} + {\tt crpx},
\label{eqn:rad2pix}
\end{eqnarray}
where {\tt pix} and {\tt cprx} are the pixel locations of $(x,y)$ and 
the CEA patch center 
relative to the lower-left corner of the patch.

\begin{figure}
\centerline{\includegraphics[width=1.0\textwidth,clip=]{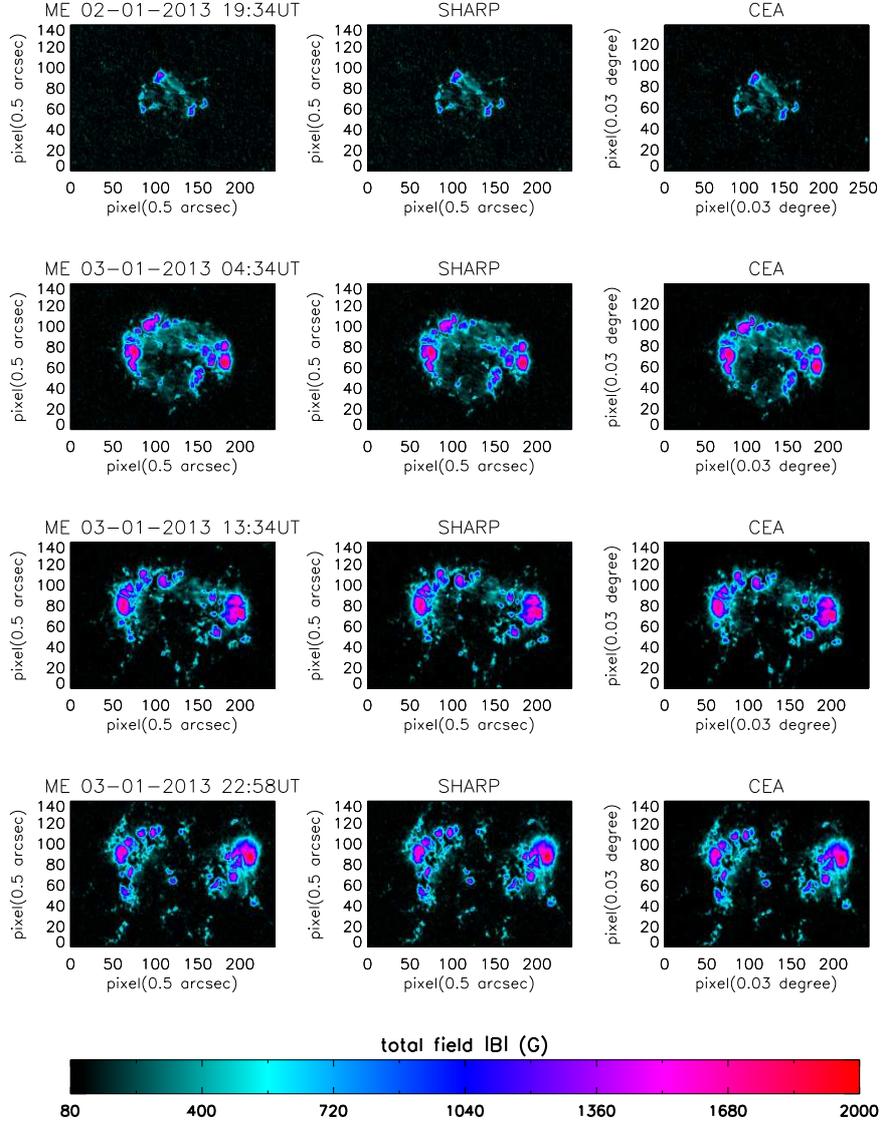}}
\caption{
Selected total field ($|B|$) images to illustrate 
the accuracy of the alignment between
the cuts from HMI full-disk vector magnetograms (ME, left column),
the Spaceweather HMI Active Region Patches (SHARP; middle),
and the Cylindrical Equal Area mapped patches (CEA; right).
The images in the same row correspond to the same observation time,
which is indicated above the left panel.
The length of each pixel of ME and SHARP is 0.5 arcsec,
and that of CEA is $0.03^{\circ}$, as indicated along respective axis.
}
\label{fig:ME+SHARP+CEA}
\end{figure}

Figure~\ref{fig:ME+SHARP+CEA} shows the comparison of
some selected ME (left), 
and corresponding SHARP (middle) and CEA (right) images.
The ME images here have been corrected for the $180^\circ$ rotation
for easier comparison with the SHARP and CEA images.
It can be seen from the figure that
the regions cut in the ME, SHARP, and CEA images are consistent.

\subsection{Horizontal Motion}
\begin{figure}
\centerline{\includegraphics[width=1.0\textwidth,clip=]{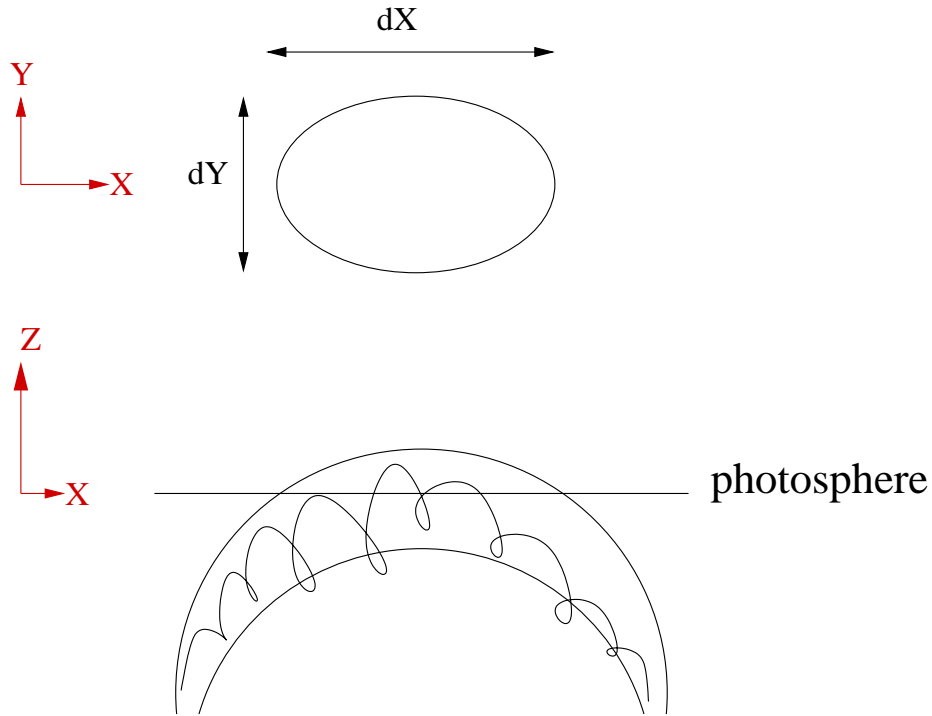}}
\caption{
A cartoon illustrating the intersection of
an emerging flux rope with the photosphere,
which is represented by the horizontal line in the lower plot.
$dX$ represents the length of the cross section of the emerged flux rope,
and corresponds to the separation of the two polarities.
$dY$ represents the width of the cross section,
and corresponds to the extension of the emerging flux region.
}
\label{fig:EFR}
\end{figure}

The geometry of the emerging flux region was measured considering
the separation of the two polarities
and the extension of the region in the direction perpendicular
to the separation of the two polarities.
The extension is used as a proxy for the width of
the flux tube (see the sketch in Figure~\ref{fig:EFR}).
For this selected EAR,
the separation of the two polarities is mostly in the East-West (X) direction
parallel to the Equator,
and the extension of the region is mainly in the South-North (Y) direction.
Therefore, for the rest of the paper,
the separation and extension will be represented by $dX$ and $dY$,
respectively.

The temporal change of $dX$ was determined from an X-t plot
created by averaging the total field $|B|(X,Y,t)$ 
over the Y direction.
Since our interest is the main part of the EAR,
which we loosely define as the region showing rapid flux emergence
and/or with strong magnetic field,
the peripheral area was not included in the averaging
to avoid contamination:
\begin{eqnarray} 
|B|(X,t) = \frac{1}{N_{Y_2}-N_{Y_1}+1}  \sum_{j=N_{Y_1}}^{N_{Y_2}} |B|(X, Y_j, t),
\end{eqnarray}
where $N_{Y_1}$ ($N_{Y_2}$) is the pixel location of 
the lower (upper) limit of the averaging range in Y,
and $N_{Y_2}-N_{Y_1}+1$ is the total number of pixels
over which $|B|$ is averaged.
Similarly, $dY$ as a function of time was determined from an Y-t plot
created by averaging $|B|(X,Y,t)$ 
over the X direction of the main part of the EAR:
\begin{eqnarray} 
|B|(Y, t) = \frac{1}{N_{X_2}-N_{X_1}+1}  \sum_{i=N_{X_1}}^{N_{X_2}} |B|(X_i, Y, t),
\end{eqnarray}
where $N_{X_1}$ ($N_{X_2}$) is the pixel address of the left (right) end of 
the averaging range in X,
and $N_{X_2}-N_{X_1}+1$ is the total number of pixels over which
$|B|$ is averaged.
$dX(t)$ and $dY(t)$ were then determined by manually tracing the edges of
the area with high intensity of $|B|$ in the X-t and Y-t images, respectively.
The tracing was repeated at least five times.
The level of the scattering of the five tracing results 
is used as a visual measure of the uncertainties in the
$dX(t)$ and $dY(t)$, and their subsequently derived quantities.

\begin{figure}
\centerline{\includegraphics[width=1.0\textwidth,clip=]{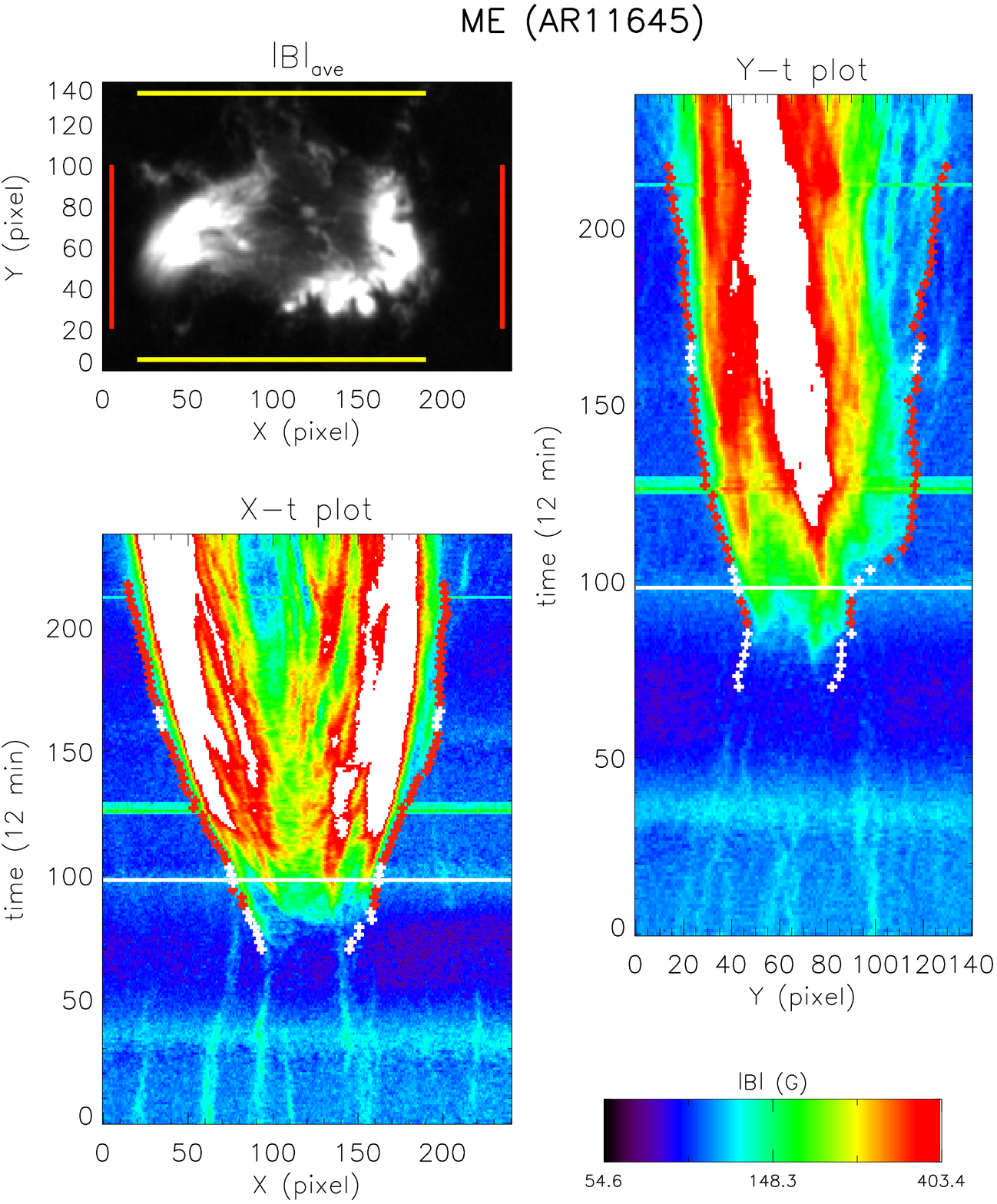}}
\caption{
Upper left, lower left, and upper right panels are
$|B|_{\rm ave}$, X-t and Y-t plots
created by averaging the total field data cube, $|B|(X,Y,t)$,
over $t$, $Y$ and $X$ dimensions, respectively.
The red vertical and yellow horizontal lines in the upper left panel
indicate the ranges over which the averages are done in $Y$ and in $X$, 
respectively.
The average range in $t$ is the entire observation time.
The data cube is the HMI full-disk vector magnetogram product ME.
Due to a combination of satellite and instrument roll angles, 
$|B|_{\rm ave}$ is rotated $180^\circ$ with respect to the real image,
and the left and right of the X-t and Y-t plots are switched 
so that left should be right
and {\it vice versa}.
In both X-t and Y-t plots, the vertical direction is time
(12 min {\it per} pixel), 
and the horizontal direction is length (0.5 arcsec {\it per} pixel).
The magnitude of the total field in the X-t and Y-t plots is shown
in the color bar in the lower right corner.
The white regions in the X-t and Y-t plots are the regions 
with field strength higher than 
our plotting maximum $403.4$~G.
$|B|_{\rm ave}$ in the top left panel provides an overall idea
of the emerging active region, 
and therefore the exact field strength is not shown.
The red and white thick crosses in the X-t and Y-t plots are 
the manually traced edges of
the emerging flux region (see the text for the difference between them).
The white horizontal line marks the starting point of 
the HMI Active Region Patch data product (SHARP and CEA).
}
\label{fig:XtYt_ME}
\end{figure}
\begin{figure}
\centerline{\includegraphics[width=1.0\textwidth,clip=]{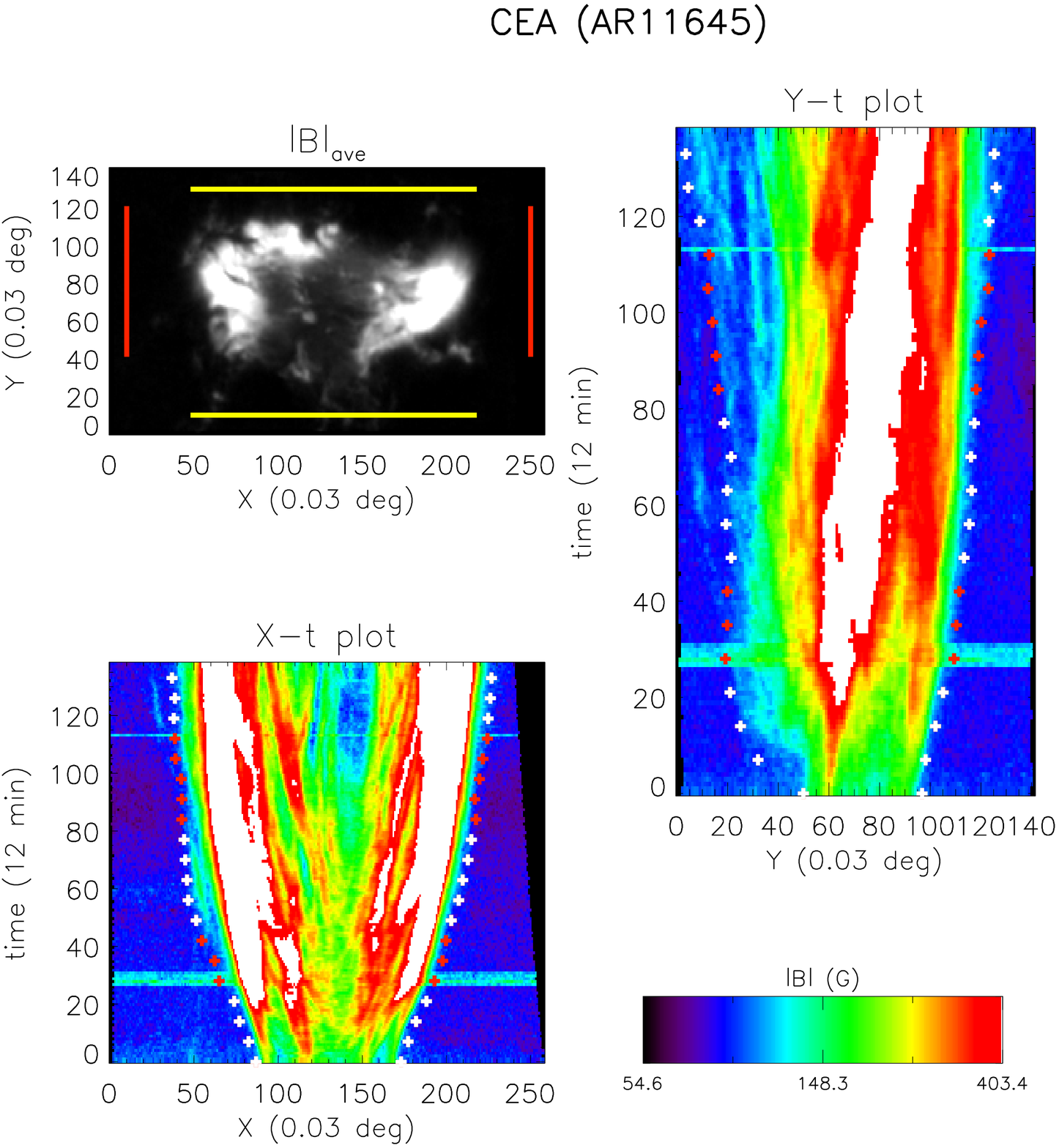}}
\caption{
Upper left, lower left, and upper right panels are
$|B|_{\rm ave}$, X-t and Y-t plots
created by averaging the total field data cube, $|B|(X,Y,t)$,
over $t$, $Y$ and $X$ dimensions, respectively.
The red vertical and yellow horizontal lines in the upper left panel
indicate the ranges over which the averages are done in $Y$ and in $X$,
respectively.
The average range in $t$ is the entire observation time.
The data cube is the HMI Active Region Patch product CEA,
and the plots are in the same orientation as the real image.
In both X-t and Y-t plots, the vertical direction is time
(12 min {\it per} pixel), 
and the horizontal direction is length ($0.03^{\circ}$ {\it per} pixel).
The magnitude of the total field in the X-t and Y-t plots is shown
in the color bar in the lower right corner.
The white regions in the X-t and Y-t plots
are the regions with field strength higher than our
plotting maximum $403.4$~G.
$|B|_{\rm ave}$ in the top left panel provides an overall idea
of the emerging active region,  
and therefore the exact field strength is not shown.
The red and white thick crosses in the X-t and Y-t plots are
the manually traced edges of
the emerging flux region (see the text for the difference between them).
}
\label{fig:XtYt_CEA}
\end{figure}

X-t and Y-t plots together with one tracing are shown
in Figure~\ref{fig:XtYt_ME} for the ME data and
in Figure~\ref{fig:XtYt_CEA} for the CEA data.
In both figures,
the upper left panel is the total $|B|$ averaged over time
($|B|_{\rm ave}=\int |B|(X,Y,t) dt/T$, 
where $T$ is the temporal length of the observation),
which illustrates the overall shape of the EAR.
The two red vertical lines in this panel 
mark the range in Y over which the average is done to create the X-t plot, 
which is placed in the lower left,
and the two yellow horizontal lines mark the range in X over which
the average is done to create
the Y-t plot, shown in the upper right corner.
The crosses in the X-t and Y-t plots mark 
the results of one manual edge tracing.

Note that the HMI instrument rotation was not corrected
in Figure~\ref{fig:XtYt_ME}.
Therefore, the $|B|_{\rm ave}$ image is rotated approximately 180$^{\circ}$
with respect to the real image,
and, consequently, the left and right of the X-t and Y-t plots are 
switched so that the left should be right and {\it vice versa}.
The white solid lines in the X-t and Y-t plots of Figure~\ref{fig:XtYt_ME}
mark the starting point of the CEA data.
The two horizontal stripes in the X-t and Y-t plots 
(at $t\approx 130$ and $\ge 200$)
were caused by an
unknown sudden change of the image intensity scale during the observation.
To check whether the determined edges encompass the main part of
the emerging flux region,
several selected ME and CEA images,
marked by the white crosses in the X-t and Y-t plots,
were plotted in Figure~\ref{fig:dXdY_ME} for ME data
and in Figure~\ref{fig:dXdY_CEA} for CEA data.
The overplotted white boxes are the manually determined boundaries.
The ME images are, as noted earlier, 180$^{\circ}$ rotated.

The images in the first two rows of Figure~\ref{fig:dXdY_ME} show that
the magnetic flux change is very dynamic during
the earliest stage of the emergence:
small fluxes can randomly appear at random locations with arbitrary orientation,
and then quickly disappear, dissipate, move away, change orientation
or converge.
This dynamic appearance has little resemblance with the simplistic picture
of a single emerging flux rope, as sketched in Figure~\ref{fig:EFR},
and causes uncertainties 
in the determination of the boundary of the flux emergence region.
Despite such difficulty,
the white boxes do cover the main part of the emerging region,
but do not always include all of the fluxes.
We found that
the excluded magnetic regions are often preexisting flux regions
or flux regions that have moved away and/or dissipated quickly.
Since our focus is the main part of the flux rope,
these outlying fluxes should not be included.
After $dX$ and $dY$ were determined,
$(dX\cdot dY)$,
which will be referred to as $(dXdY)$ hereinafter,
was used as a measure of area of the cross section of
the emerging magnetic flux rope.
The changing rate of all these parameters was computed by taking the time derivatives:
$V_X \equiv \partial dX/\partial t$,
$V_Y \equiv \partial dY/\partial t$, and 
$V_{XY} \equiv \partial (dXdY)/\partial t$.

\subsection{Vertical Motion}
\citet{Schuessler1979AA} was the first to
derive an analytical expression for the magnetic buoyant force of
a horizontal cylindrical magnetic tube with an arbitrary radius,
under the simplification of constant temperature,
uniform magnetic field, constant gravity, and constant pressure scale-height
across the cross section, which is
\begin{eqnarray}
\widetilde{F}_B &=& \frac{1}{4} B_0^2 \cdot a \cdot \exp(-a/\Lambda) \cdot
                 I_1(a/\Lambda),
\label{eqn:F_B1}
\end{eqnarray}
where
$\widetilde{F}_B$ is the magnetic buoyant force {\it per} unit length,
$a$ is the tube radius,
$B_0$ and $\Lambda$ are the magnetic field and
pressure scale-height of the cross section of the tube, and
$I_1$ is the modified Bessel function of order 1.
In the limit of $a/\Lambda \ll 1$ (thin flux-tube approximation) and of
$a/\Lambda \gg 1$ (thick flux-tube approximation),
$\widetilde{F}_B$ simplifies to the following: 
\begin{eqnarray}
\widetilde{F}_B &\approx& \frac{B_0^2 \cdot a^2}{8\Lambda} 
\;\;\;\;\;  (a/\Lambda \ll 1) \mbox{ and}
\label{eqn:F_Bthin} \\
\widetilde{F}_B &\approx& \frac{B_0^2}{4 (2\pi)^{1/2}}(a \cdot \Lambda)^{1/2}
\;\;\;\;\;  (a/\Lambda \gg 1).
\label{eqn:F_Bthick} 
\end{eqnarray}
Using Equations~(\ref{eqn:F_Bthin}) and (\ref{eqn:F_Bthick}) and
assuming that the upward buoyant force is balanced by
a downward drag force of the following form:
\begin{eqnarray}
\widetilde{F}_d &=& \frac{1}{2}\rho_e v^2 C_d \cdot 2a,
\label{eqn:F_d}
\end{eqnarray}
where $\widetilde{F}_d$ is the drag force {\it per} unit length,
$\rho_e$ is the ambient gas density,
$v$ is the velocity, and $C_d$ is drag coefficient,
\citet{CW1987SoPh} derived the buoyant velocities for the
thin and thick flux-tube approximations, as:
\begin{eqnarray}
V_{\rm bn} &=& 
V_A \left(\frac{\pi}{2C_d}\right)^{1/2} \left(\frac{a}{\Lambda}\right)^{1/2}
\;\;\;\;\; (a \ll \Lambda) \mbox{ and}
\label{eqn:Vbn} \\
V_{\rm bk} &=&
V_A \left(\frac{\pi}{2C_d^2}\right)^{1/4}\left(\frac{\Lambda}{a}\right)^{1/4}
\;\;\;\;\; (a \gg \Lambda),
\label{eqn:Vbk}
\end{eqnarray}
where $V_A=B_0/\sqrt{4\pi\rho_e}$ is the Alfv\'en speed,
and $V_{\rm bn}$ and $V_{\rm bk}$ denote 
the thin and thick flux-tube buoyant velocities,
respectively.
To derive the buoyant velocities at the surface layer,
the parameters $\rho_e$, $\Lambda$ and $C_d$ 
are set to their photospheric values:
\begin{eqnarray}
\rho_e &=& 3.2 \times 10^{-7} \; {\rm g}\;{\rm cm}^{-3}, \\
\Lambda &=& 200 \; {\rm km} = 2 \times 10^7 \; {\rm cm}, \\
C_d &=& 1 \mbox{   (dimensionless)}.
\end{eqnarray}
Equations (\ref{eqn:Vbn}) and (\ref{eqn:Vbk})
show that the buoyant velocities are mainly dependent on
the magnetic field strength and radius of the flux tube, 
specifically, $V_{\rm bn} \propto B_0 a^{1/2}$ and
$V_{\rm bk} \propto B_0 a^{-1/4}$.
Since our analysis is based on the hypothesis that
the entire emerging active region is formed by
a single flux tube,
we used $dY/2$ as a measure of the radius,
and the magnetic field strength was set to be
the average magnetic field of the emerging flux region.

\begin{figure}
\centerline{\includegraphics[width=1.0\textwidth,clip=]{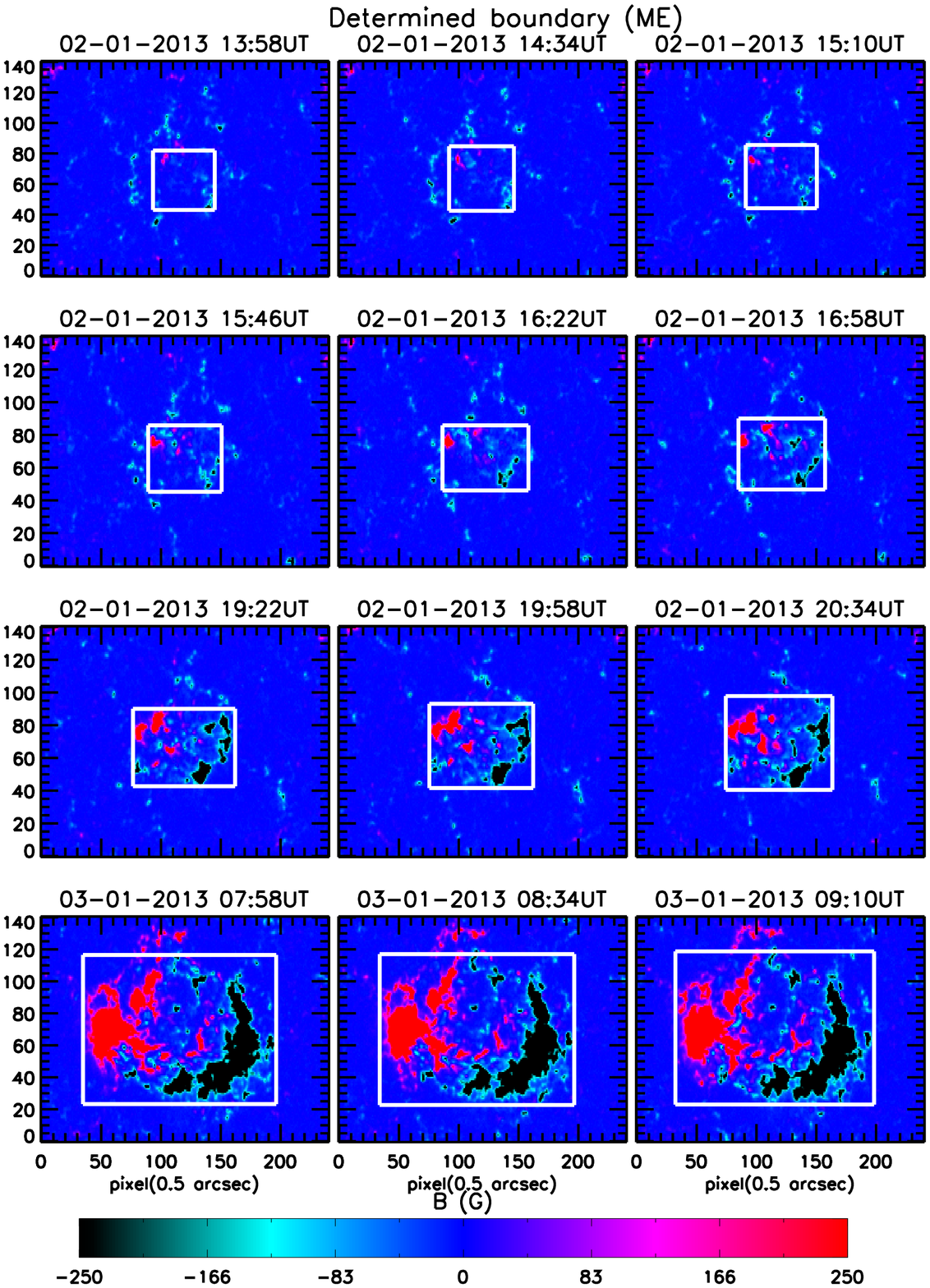}}
\caption{
Selected images to check the accuracy of the EAR boundaries 
determined from the X-t and Y-t plots of the
HMI vector magnetogram product ME (see Figure~\ref{fig:XtYt_ME}).
The images plotted here correspond to the white crosses 
in Figure~\ref{fig:XtYt_ME}.
The white boxes in each panel are the determined boundaries. 
The exact observation time is indicated above each panel.
The pixel size in both X and Y directions is 0.5 arcsec,
as indicated in the X-axis labels of the bottom row.
We show the line-of-sight component of the field,
instead of the total field,
to distinguish the two polarities
}
\label{fig:dXdY_ME}
\end{figure}

\begin{figure}
\centerline{\includegraphics[width=1.0\textwidth,clip=]{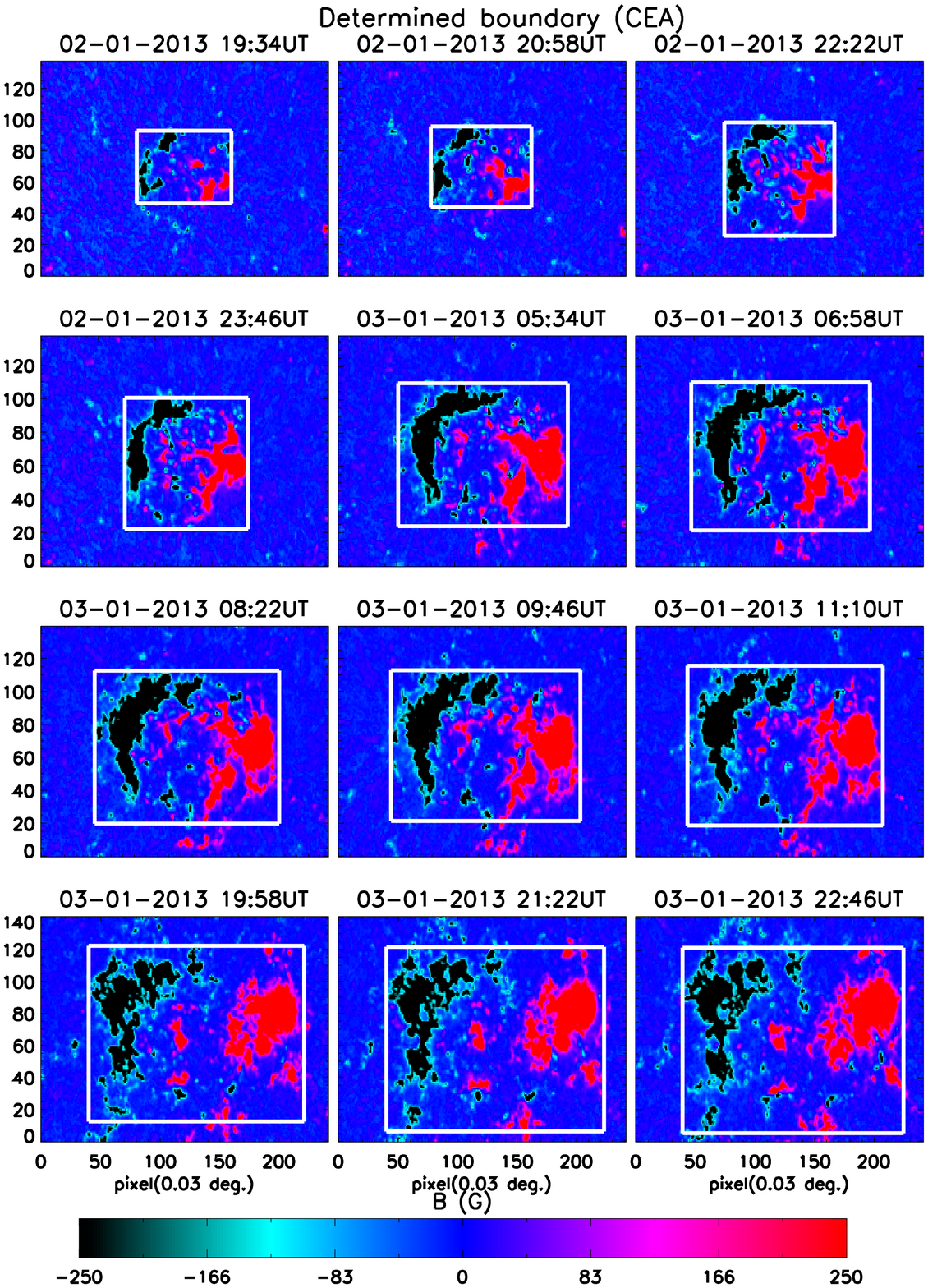}}
\caption{
Selected images to check the accuracy of the EAR boundaries
determined from the X-t and Y-t plots of the HMI vector magnetogram product
CEA (see Figure~\ref{fig:XtYt_CEA}).
The images plotted here correspond to the white crosses in Figure~\ref{fig:XtYt_CEA}.
The white boxes in each panel indicate the determined boundaries.
The observation time is indicated above each panel.
The pixel size in both X and Y direction is 0.03 degree,
as indicated in the X-axis labels of the last row.
We show the radial component of the field,
instead of the total field,
to distinguish the two polarities
}
\label{fig:dXdY_CEA}
\end{figure}

As shown in Figures~\ref{fig:dXdY_ME} and \ref{fig:dXdY_CEA},
a large portion of the area bounded by $dX$ and $dY$ have
very weak magnetic field strength.
To calculate a meaningful average magnetic field of 
the emerging flux region,
the area with weak background field should not be included.
To distinguish the emerging flux region from the background field,
we applied a threshold to filter out weak-field regions,
that is, the regions with $|B|$ less than a threshold value, $|B|_{\rm thr}$,
were neglected.
Because the field strength of the emerging flux changes significantly
from the earliest stages of emergence to its maximum value,
the threshold should not be fixed but should be adjusted
according to the stage of the emergence.
The threshold should be sufficiently high to exclude most of the noise
but also sufficiently low such that 
the initial emerging fluxes are not overlooked.
Since an emerging magnetic flux structure is expected to 
be more concentrated than the background field and
should first appear at the photosphere as horizontal,
the threshold $|B|_{\rm thr}$ was chosen to be the average field strength of
the inclined fields within $dX$ and $dY$.
``Inclined'' field in this study
is defined as the field vector with
inclination angle $\theta_B$ between $50^\circ$ and $130^\circ$,
following the criteria in \citet{centeno2012ApJ}.
Therefore, 
\begin{eqnarray}
|B|_{\rm thr} &=& \frac{\sum |B|^{\rm inclined}}{N_{50^{\circ}<\theta_B<130^{\circ}}},
\end{eqnarray}
where $\sum |B|^{\rm inclined}$ is the sum of $|B|$ of the
inclined field vectors,
and $N_{50^{\circ}<\theta_B<130^{\circ}}$ is the total number of pixels 
of the inclined fields.
We point out that the inclination angle of the ME data is measured relative 
to LOS
while that of the CEA data is measured relative to the radial direction.
Therefore,
if the projection effect needs to be considered,
that is, if the LOS is far from the local vertical direction,
the results of both should differ.

After the threshold was determined,
the average field was calculated in two different ways:
1) averaging over all strong fields
({\it i.e.}, $|B|>|B|_{\rm thr}$):
\begin{eqnarray}
\langle|B|^{\rm all}\rangle &=& 
\frac{\sum_{|B|>|B|_{\rm thr}} |B|}{N^{\rm all}},
\end{eqnarray}
and 2) averaging over the strong field where the field vector is inclined:
\begin{eqnarray}
\langle|B|^{\rm inclined}\rangle &=& 
\frac{\sum_{|B|^{\rm inclined}>|B|_{\rm thr}} |B|^{\rm inclined}}{N^{\rm inclined}},
\end{eqnarray}
where $N^{\rm all} \equiv N_{|B|>|B|_{\rm thr}}$ is
the total number of pixels with strong field values,
and $N^{\rm inclined} \equiv N_{|B|^{\rm inclined}>|B|_{\rm thr}}$
is the total number of pixels with strong field values 
for which the field vector is inclined.
Note that the difference $N^{\rm all}-N^{\rm inclined}$ is simply
the number of pixels with strong fields and
vertically oriented field vectors.
For the rest of this article,
the average field and associated quantities
of the former will be referred to as ``all-direction'' and labeled
with the superscript ``all'',
while those of the latter will be referred to as ``inclined-field'' and labeled
with the superscript ``inclined''.

\begin{figure}
\centerline{\includegraphics[width=1.0\textwidth,clip=]{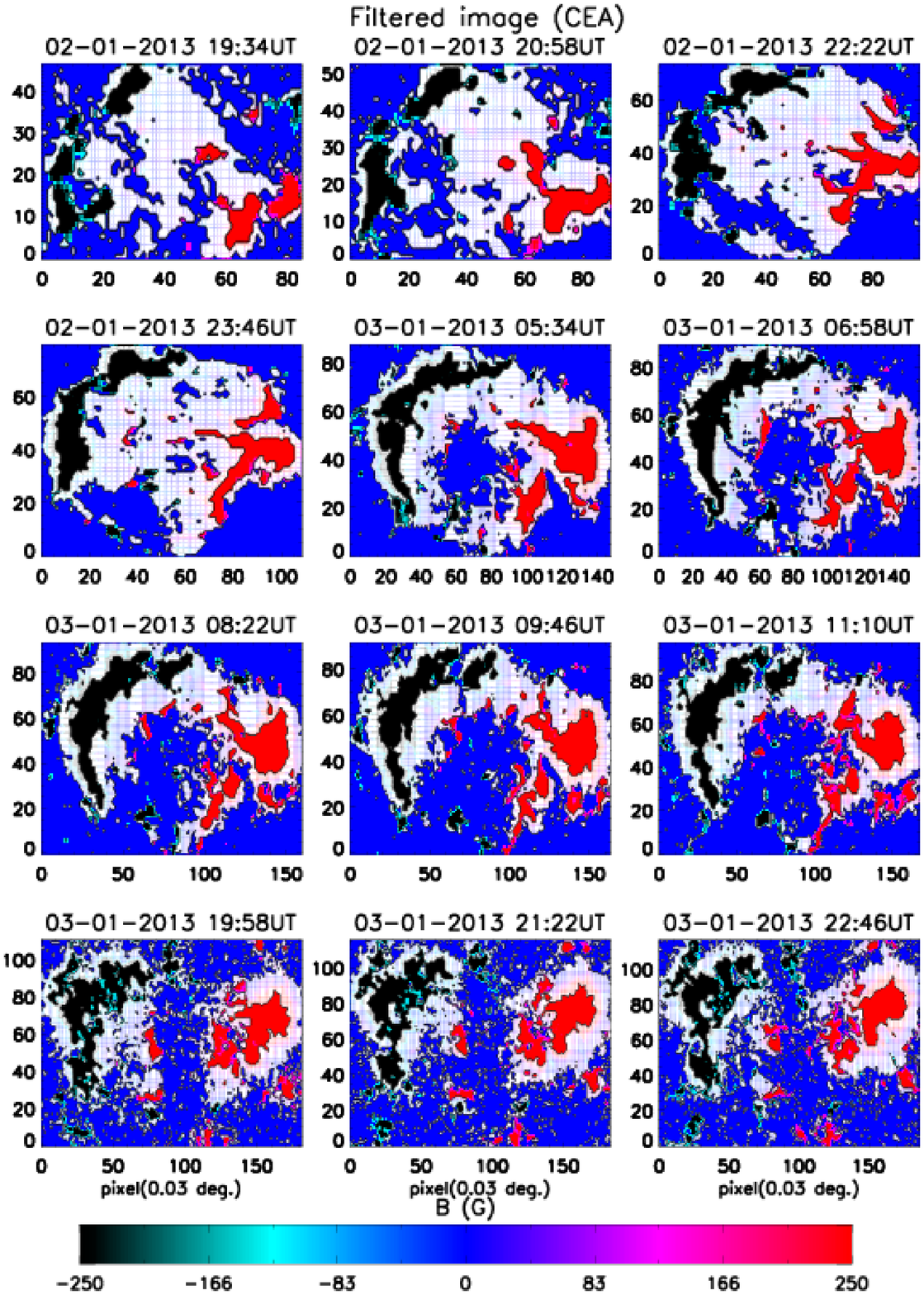}}
\caption{
Results after filtering the weak-field regions.
The total field lower than the threshold ($|B|<|B|_{\rm thr}$)  
is set to zero.
These images correspond to the regions within the white boxes in
Figure~\ref{fig:dXdY_CEA}.
The observation times are as indicated above the corresponding panels.
The pixel size in both X and Y directions is 0.03 degree,
as indicated in the X-axis labels in the last row.
The areas with inclined field vectors
($50^\circ<\theta_B<130^\circ$)
are painted in white to distinguish them
from those with more vertically oriented field vectors.
The quantity shown in the images is the radial component of
the field, instead of the total field,
to distinguish the two polarities.
}
\label{fig:Bavmin_CEA}
\end{figure}

The results using ME and CEA data after filtering are shown in 
Figure~\ref{fig:Bavmin_ME} and Figure~\ref{fig:Bavmin_CEA}, respectively.
The plotted regions are inside the white boxes shown
in Figures~\ref{fig:dXdY_ME} and \ref{fig:dXdY_CEA}.
The fields lower than $|B|_{\rm thr}$ were 
set to zero (blue).
The regions with non-zero values and inclined fields are shown in white to
distinguish them from the regions with more vertically oriented field.
Comparing Figures~\ref{fig:dXdY_ME} and \ref{fig:dXdY_CEA} with
Figures~\ref{fig:Bavmin_ME} and \ref{fig:Bavmin_CEA},
we find that the first four images are still very noisy after the filtering,
indicating that $|B|_{\rm thr}$ during the early stage of the flux emergence
is comparable to the noise level,
causing some stronger random noise to be included as part of the emerging flux region.
However,
since
the only relevant output from this filtering process is the average total-field strength,
such imperfect filtering should not introduce significant errors to
$\langle|B|^{\rm all}\rangle$ and $\langle|B|^{\rm inclined}\rangle$
and the subsequently computed $V_{\rm bk}$ and $V_{\rm bn}$,
as long as the field strength of the included noise is similar to that in
the emerging flux region.
Our second observation is that
the field vectors were predominantly inclined during
the early stage of emergence,
but became more vertically oriented later.
This is consistent with the expected characteristics 
of a magnetic flux tube rising through the photosphere,
and has been widely reported in both observations and theoretical models
\citep[{\it e.g.},][]{CF1989ApJ,chou1993ASPC,Caligari_etal1995ApJ,lsmp1998AA,Fan2004ASPC,centeno2012ApJ}.
Our third observation is that
the two polarities were connected by inclined fields
during most of the emerging process,
but gradually they broke up to become two separated regions,
each of which is surrounded by a ring of inclined fields
({\it cf.} last row of Figure~\ref{fig:Bavmin_CEA}).
These observations are consistent with general properties
of emerging active regions.

After $\langle|B|^{\rm all}\rangle$ and $\langle|B|^{\rm inclined}\rangle$ were determined,
we used $dY/2$ as a proxy for the flux tube radius
and derived the buoyant velocities for thick and thin flux-tube approximation
({\it i.e.,} $V_{\rm bk}$ and $V_{\rm bn}$).
We also computed the changing rates of $\langle|B|^{\rm all}\rangle$, 
$\langle|B|^{\rm inclined}\rangle$, $V_{\rm bk}$, and $V_{\rm bn}$.

\section{Results and Discussion} \label{sec:result}
\begin{figure}
\centerline{\includegraphics[width=1.0\textwidth,clip=]{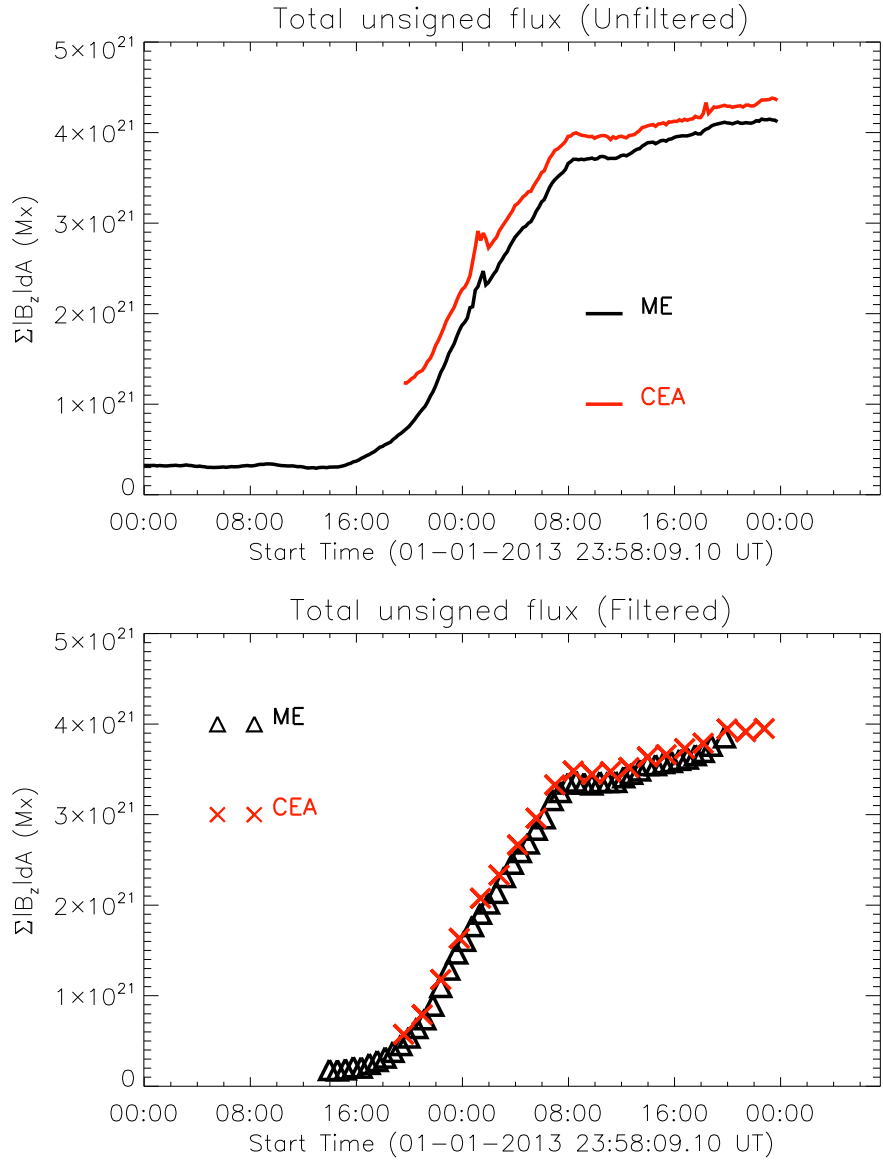}}
\caption{
Temporal profile of the total unsigned flux
of the analyzed region, $F(t)=\sum |B_z|dA$.
The results computed from the ME and CEA magnetograms are shown in black and red,
respectively.
The upper panel shows the result of the original data without filtering,
and the lower panel shows the result after the weak field,
$|B|<B_{\rm thr}$, has been filtered.
}
\label{fig:flx}
\end{figure}

The temporal profiles of the total unsigned flux in the entire region cut
from HMI full-disk magnetograms
are plotted in Figure~\ref{fig:flx} to show how the total flux
was changing during the flux emergence.
The upper panel shows the results computed from the original
ME (black) and CEA (red) data without any filtering.
The lower panel shows the results of the filtered data,
that is, only $|B|>|B|_{\rm thr}$ was included in the summation.
Both plots show
that the total flux began to increase
around 2 January 2013 at 16:00~UT.
The main emergence lasted for approximately 16 hours
until around 3 January 2013 at 08:00~UT, after which
the increasing rate became lower.
Comparison between the two plots shows that
the unfiltered results contain
two spikes (at $\approx$~01:00~UT and 18:00~UT on 3 January 2013)
and a gap between the ME and CEA profiles.
Both disappeared in the filtered result in the lower panel.
This indicates that both features are likely caused by
the weak-field noise.

\begin{figure}
\centerline{\includegraphics[width=1.0\textwidth,clip=]{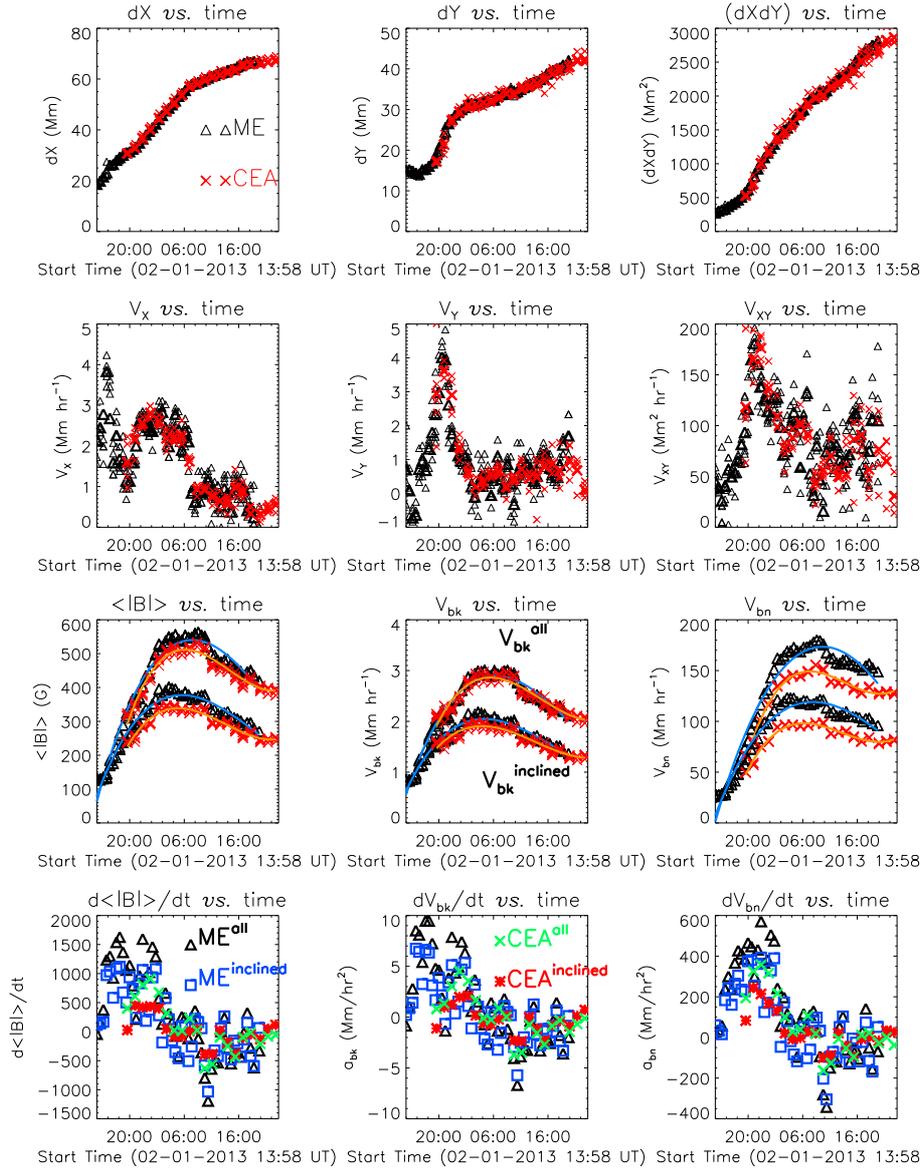}}
\caption{
Measured quantities {\it vs.} time.
The upper two rows show the quantities associated with the horizontal motions
($dX$, $dY$, $(dXdY)$, and their temporal derivatives),
and the lower two rows show those associated with the vertical motions
($\langle|B|\rangle$, $V_{\rm bk}$, $V_{\rm bn}$,
and their time derivatives).
$\langle|B|\rangle$, $V_{\rm bk}$, and $V_{\rm bn}$
in the plots are the notations for
the all-direction and inclined-field quantities.
In the upper three rows, the red crosses and black triangles represent
the results of CEA and ME data, respectively.
In the upper two rows, the thinner symbols represent 
the results of individual edge-tracing of the X-t and Y-t plots
to provide a visual measure of the uncertainties of 
the plotted quantities,
and their averages are plotted as thicker ones
to show their temporal profiles.
In the third row, 
the higher curves correspond to the all-direction results,
and the lower ones the inclined-field results.
The solid lines are the results of a third degree polynomial fitting
to the data.
In the last row, black triangles, blue squares, green crosses and red stars
represent ME$^{\rm all}$, ME$^{\rm inclined}$, CEA$^{\rm all}$ and CEA$^{\rm inclined}$,
as indicated
}
\label{fig:kin1}
\end{figure}

In Figure~\ref{fig:kin1}, 
the quantities associated with the horizontal motion
($dX$, $dY$, $(dXdY)$, and their temporal derivatives) of the EAR
are plotted in the upper two rows
and those associated with the vertical motion 
($\langle|B|^{\rm all}\rangle$, $\langle|B|^{\rm inclined}\rangle$, 
$V_{\rm bk}$, $V_{\rm bn}$, 
and their temporal derivatives) in the lower two.
The plots in the upper two rows show that
the horizontal motion determined from ME (black) and CEA (red) data
are consistent with each other.
The time profiles of $dX$ and $V_X$ show
that the separation of the two polarities was fastest 
($\approx 4.0$~Mm~hr$^{-1}$)
during the initial emergence,
as indicated by the steep slope in $dX$ {\it vs.} time and the peak in 
$V_X$ {\it vs.} time
at $\approx$~16:00~UT on 2 January 2013.
The two polarities continued to separate at a lower speed afterward.
The two plateaus in $V_X$ {\it vs.} time indicate that
the separation speed decreased twice,
$\approx 3$~Mm hr$^{-1}$ at around 01:00~UT on 3 January 2013 and
$\approx 1$~Mm hr$^{-1}$ at around 11:00~UT on 3 January 2013.
Our result is qualitatively similar to those found in
earlier studies
\citep[{\it e.g.},][]{HM1973SoPh,CW1987SoPh,strous_etal1996AA,shimizu_etal2002ApJ},
which showed that the separation of the opposite polarities is
fastest at the beginning of the emergence
but becomes slower later.
The exact values of the separation speeds, however,
differ among different studies.

The extension, $dY$, did not begin to increase until
after $\approx$~18:00~UT on 2 January 2013,
and reached its peak rate of $\approx 4.0$~Mm hr$^{-1}$
in about three hours at $\approx$~21:00~UT on 2 January 2013.
The increase of $dY$ began to slow down after $\approx$~23:00~UT on 2 January 2013,
and stayed at an almost constant rate of $\approx 1$~Mm hr$^{-1}$
at the end of our observation.
It is interesting to note that 
while $V_X$ and $V_Y$ seem to be anticorrelated
at the initial emerging phase,
the two have similar peak and teminal speeds.
One possible explanation
is that the peak and terminal speeds of both extension and separation
may simply depend on the total strength and dissipation of
the emerging magnetic flux tube,
while the detailed temporal profile of the motion in the two directions
may be influenced by more complex factors,
such as surface flows, turbulence, and others.

\begin{figure}
\centerline{\includegraphics[width=1.0\textwidth,clip=]{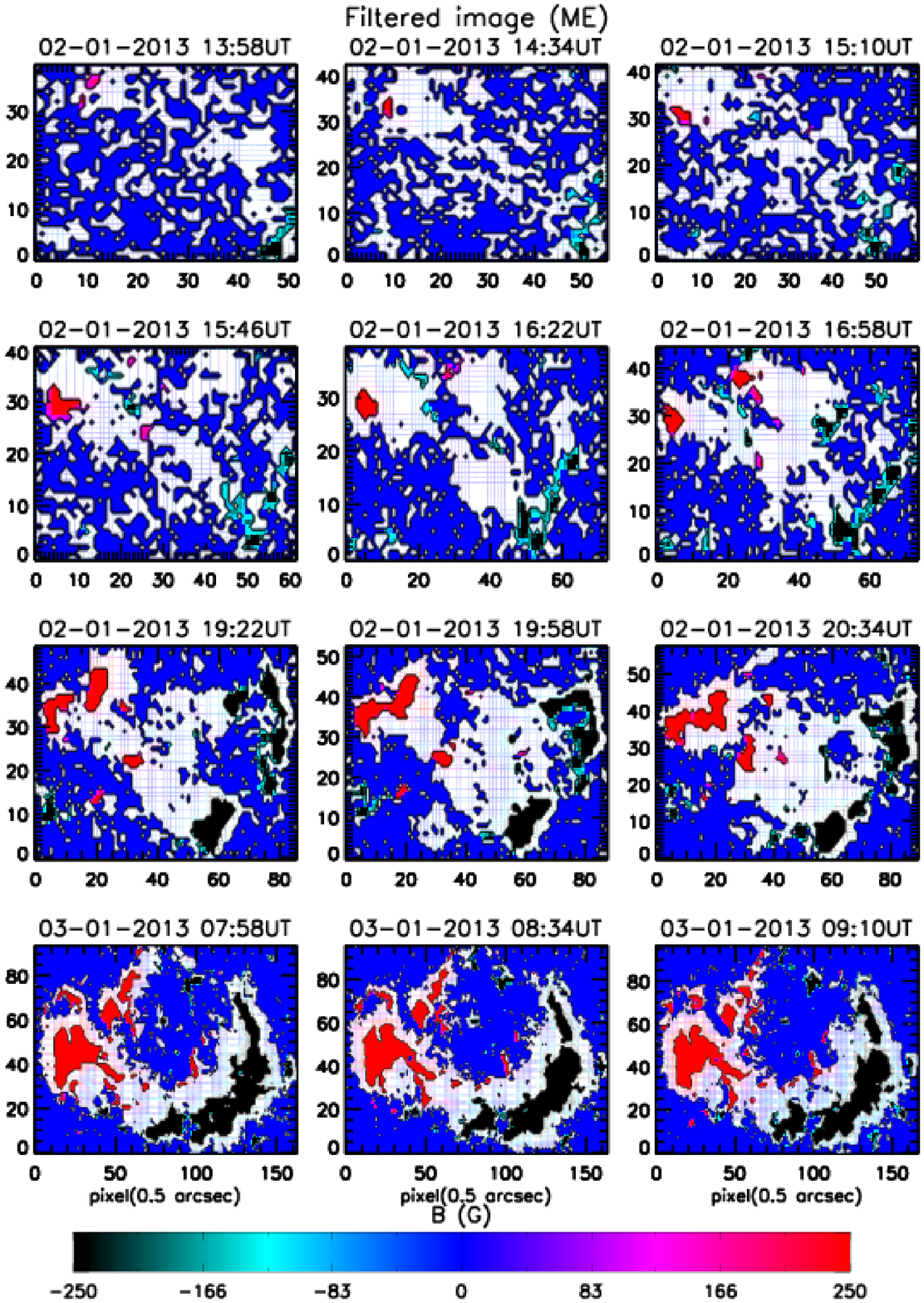}}
\caption{
Results after filtering the weak-field regions.
The total field lower than the threshold ($|B|<|B|_{\rm thr}$) 
is set to zero.
These images correspond to the regions within the white boxes in 
Figure~\ref{fig:dXdY_ME}.
The observation times are as indicated above the corresponding panels.
The pixel size in both X and Y directions is 0.5 arcsec,
as indicated in the X-axis labels in the last row.
The areas with inclined field vectors
($50^\circ<\theta_B<130^\circ$) 
are painted in white to distinguish them
from those with more vertically oriented field vectors.
The quantity shown in the images is the line-of-sight component of 
the field, instead of the total field, 
to distinguish the two polarities.
}
\label{fig:Bavmin_ME}
\end{figure}

The third row shows that
the average fields and buoyant velocities derived from
ME and CEA data differ,
with the CEA results (red) lower than the ME results (black).
The difference is especially prominent in $V_{\rm bn}$ {\it vs.} time.
Comparing the all-direction (higher) and inclined-field (lower) curves 
in this row,
it can be seen that the two were
almost the same at the beginning of the emergence,
and deviated from each other later.
The gap between the two widened
as $\langle|B|^{\rm all}\rangle$, $\langle|B|^{\rm inclined}\rangle$, 
$V_{\rm bk}$, and $V_{\rm bn}$
increased,
and gradually became nearly a constant after
these four quantities reached their respective peaks.
These curves are qualitatively consistent with the image sequence in
Figures~\ref{fig:Bavmin_ME} and \ref{fig:Bavmin_CEA}.
It can also be noticed that the all-direction curves peaked slightly later than
the inclined-field curves in all three panels.
The plot of
$V_{\rm bn}$ reveals that the buoyant velocity of 
the thin flux-tube approximation
is about 20 times larger than the horizontal velocity
$V_X$ and $V_Y$,
indicating that the thin flux-tube approximation is inappropriate.
The peak $V_{\rm bk}$ is slightly smaller than the peak in $V_X$ and $V_Y$,
but the three speeds are generally of similar magnitudes.

\citet{centeno2012ApJ} examined the relationship between the inclination angles
and the Doppler velocities of all points in two emerging active regions
that emerged in a free field environment.
The scatter plots in the paper showed that the inclined-field areas
had an upward velocity of $\le 400$~m~s$^{-1}$ ($\approx 1.5$~Mm~hr$^{-1}$),
which is similar to our computed $V_{\rm bk}^{\rm inclined}$
based on the buoyancy theory.

\citet{lsmp1998AA} reported a higher upward Doppler velocity of 
$\approx 1$~km~s$^{-1}$
for horizontal magnetic elements.
However, at least one of the regions selected by \citet{lsmp1998AA} was
located within some preexisting magnetic field,
which may have affected the rising motion of the emerging flux.

An early study by \citet{CW1987SoPh},
which used LOS magnetograms,
reported a thick flux-tube buoyant velocity that was much larger than
the separation velocity.
LOS magnetograms mainly detect the vertically oriented fields,
which usually have stronger field strength 
and occur in the later stage of the flux emergence
than the horizontally oriented fields \citep[{\it e.g.},][]{ksl2003ApJ}.
As revealed in our analysis,
the separation of the two polarities quickly slowed down to less than 1~Mm~hr$^{-1}$
after $\langle|B|^{\rm inclined}\rangle$ and $\langle|B|^{\rm all}\rangle$
reached their maximum.
In contrast, the buoyant velocities remain $\ge 1$~Mm~hr$^{-1}$ at the last point
of our observation.

In the last row of Figure~\ref{fig:kin1},
we plotted the temporal derivatives of 
$\langle|B|^{\rm all}\rangle$, $\langle|B|^{\rm inclined}\rangle$,
$V_{\rm bk}^{\rm all}$, $V_{\rm bk}^{\rm inclined}$,
$V_{\rm bn}^{\rm all}$, and $V_{\rm bn}^{\rm inclined}$.
Despite the difference between the all-direction and inclined-field quantities
in the third row,
their temporal derivatives overlap, and have a similar decreasing profiles.
In other words, the changing rates of all these quantities are similar.

\begin{figure}
\centerline{\includegraphics[width=1.0\textwidth,clip=]{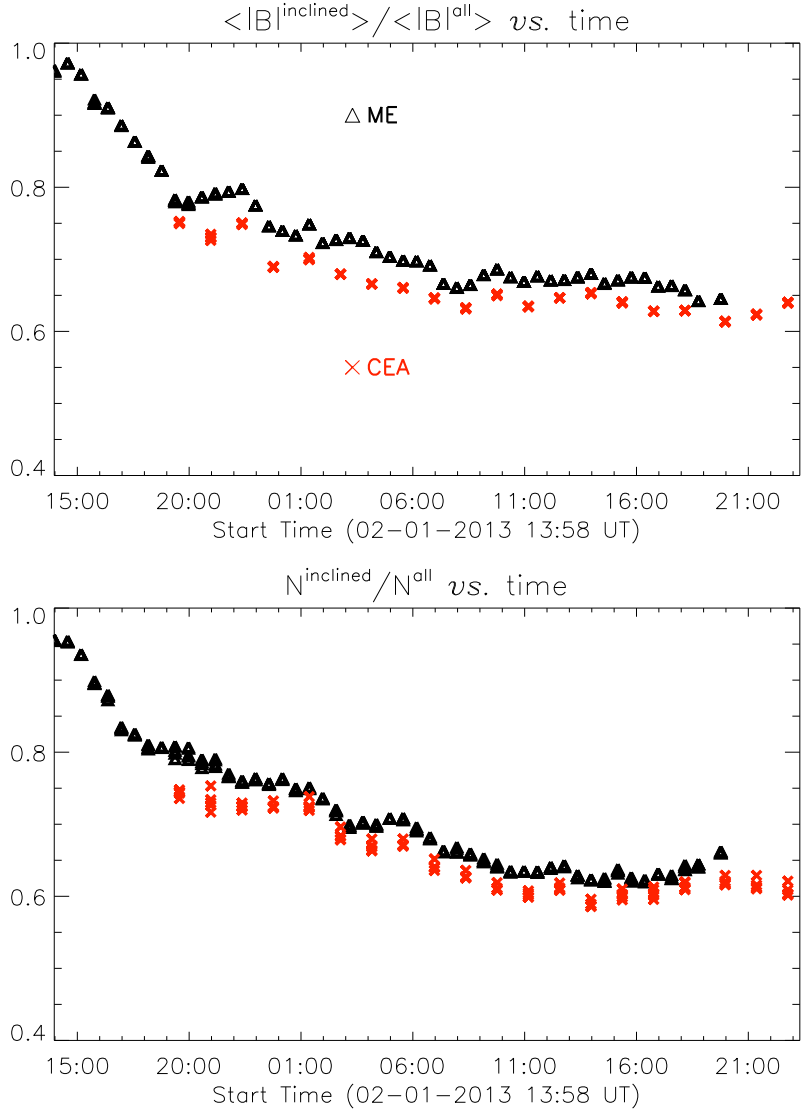}}
\caption{
The temporal profiles of 
$\langle|B|^{\rm inclined}\rangle/\langle|B|^{\rm all}\rangle$ (upper panel)
and $N^{\rm inclined}/N^{\rm all}$ (lower panel).
The black triangles and red crosses represent the results of ME and CEA data,
respectively.
At each temporal point,
there are five ME and five CEA data points,
corresponding to the individual edge-tracing results of the X-t and Y-t plots.
The level of the scattering of the data points provides 
a visual measure of the uncertainties
of $\langle|B|^{\rm inclined}\rangle/\langle|B|^{\rm all}\rangle$ and
$N^{\rm inclined}/N^{\rm all}$.
}
\label{fig:Bratio}
\end{figure}

Next, we examined how
the percentage of the inclined field in the EAR
evolves during the emerging process.
In Figure~\ref{fig:Bratio},
we plotted the time profiles of the ratio
$\langle|B|^{\rm inclined}\rangle/\langle|B|^{\rm all}\rangle$
in the upper panel,
and the ratio of the occupied area
$N^{\rm inclined}/N^{\rm all}$ in the lower panel.
The two profiles show a similar decreasing trend,
both began at $\approx 1$ and decreased to a terminal value of $\approx 0.6-0.7$.
Comparing this figure with the third row of Figure~\ref{fig:kin1},
we can see that the ratios
decreased as $\langle|B|^{\rm all}\rangle$ and 
$\langle|B|^{\rm inclined}\rangle$ increased,
and reached the terminal value at
approximately the same time as
when $\langle|B|^{\rm all}\rangle$ and $\langle|B|^{\rm inclined}\rangle$ 
reached their peaks
(at $\approx$ 06:00~UT on 3 January 2013).

To understand the plots,
we rewrite
$\langle|B|^{\rm inclined}\rangle/\langle|B|^{\rm all}\rangle$ as follows:
\begin{eqnarray}
\frac{\langle|B|^{\rm inclined}\rangle}{\langle|B|^{\rm all}\rangle} &=&
\frac{\sum |B|^{\rm inclined}/\sum |B|^{\rm all}}{N^{\rm inclined}/N^{\rm all}}.
\label{eqn:Brat}
\end{eqnarray}
Since $N^{\rm inclined} \approx N^{\rm all}$ at the beginning of the emergence,
as revealed in Figure~\ref{fig:Bavmin_ME},
$\sum |B|^{\rm all}$ must be almost equal to
$\sum |B|^{\rm inclined}$ during the initial stage of the flux emergence,
indicating that the proportion of vertically oriented fields during this time
is very small or almost negligible.
After $\langle|B|^{\rm all}\rangle$ and $\langle|B|^{\rm inclined}\rangle$
reached their maximum,
Figure~\ref{fig:Bratio} shows that
$\langle|B|^{\rm inclined}\rangle/\langle|B|^{\rm all}\rangle$ and
$N^{\rm inclined}/N^{\rm all}$ both became approximately 0.6,
leading to $\sum |B|^{\rm inclined}/\sum |B|^{\rm all} \approx 0.36$
according to Equation~(\ref{eqn:Brat}).
In other words,
while the inclined fields still cover a larger percentage of the area (60\%),
as qualitatively shown in the last row of Figure~\ref{fig:Bavmin_ME},
the sum of the inclined fields becomes only 36\% of
the sum of all-direction fields.
Since
$\sum |B|^{\rm all} = \sum |B|^{\rm inclined} + \sum |B|^{\rm vertical}$
and $N^{\rm all} = N^{\rm inclined} + N^{\rm vertical}$,
we can deduce that $\sum |B|^{\rm vertical}/\sum |B|^{\rm all} \approx 0.64$,
$N^{\rm vertical}/N^{\rm all} \approx 0.4$
and $\langle|B|^{\rm vertical}\rangle/\langle|B|^{\rm all}\rangle \approx 1.6$,
indicating that the vertically oriented fields became very concentrated
in an area about 40\% of the EAR after the growing phase of the emergence.

\begin{figure}
\centerline{\includegraphics[width=1.0\textwidth,clip=]{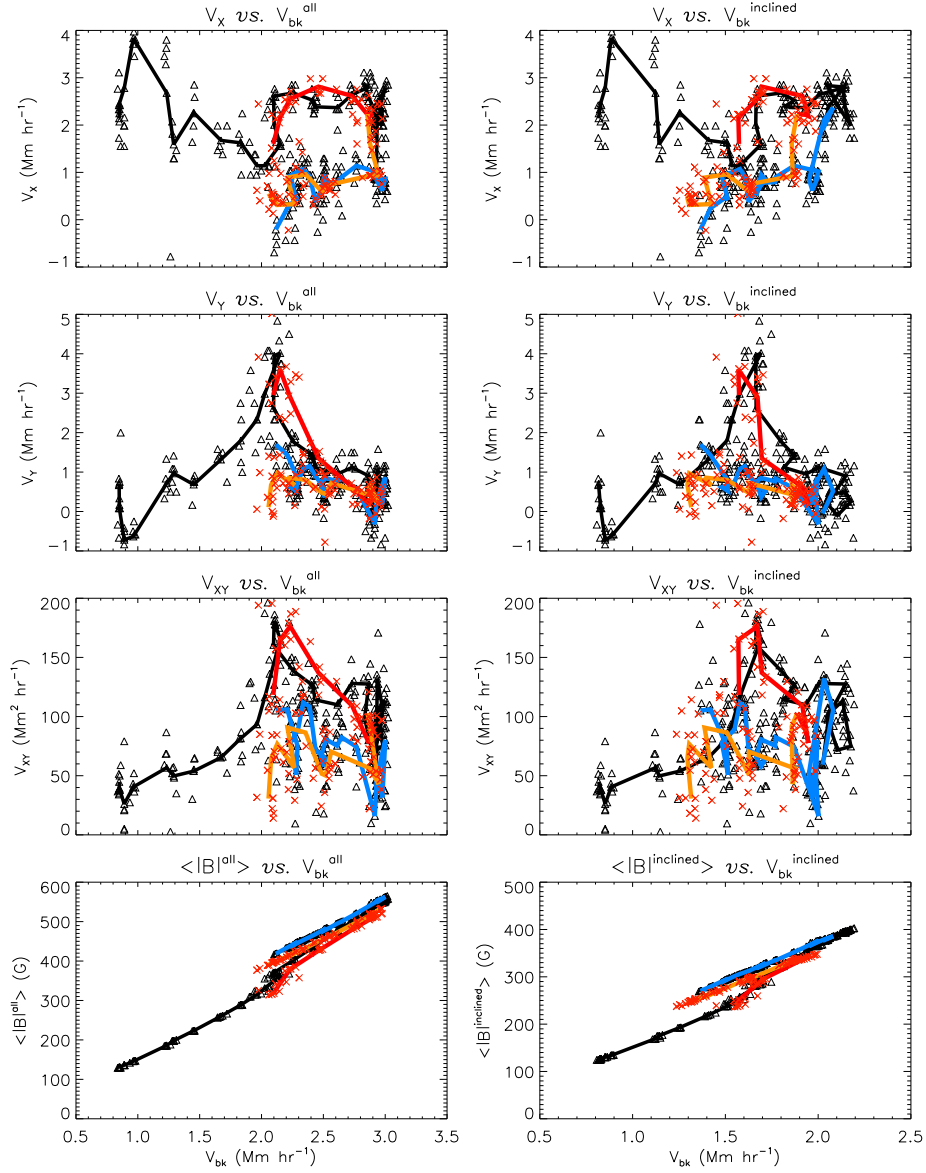}}
\caption{
From top to bottom: $V_X$, $V_Y$, $V_{XY}$ and 
$\langle|B|^{\rm all (inclined)}\rangle$ {\it vs.} 
$V_{\rm bk}^{\rm all (inclined)}$.
The all-direction and inclined-field results are in the
left and right columns, respectively.
Red crosses and black triangles correspond, respectively,
to the CEA and ME results of the individual edge tracings
to provide a visual measure of the uncertainties of the plotted quantities.
The averages of the individual tracings are joined by thick lines
to show their profiles.
The profiles before the peak
of $\langle|B|^{\rm all}\rangle$ are plotted
in black for the ME results and red for the CEA results.
After the peak of of $\langle|B|^{\rm all}\rangle$,
the ME and CEA profiles are plotted in blue and orange,
respectively.
}
\label{fig:VB2Vbk}
\end{figure}

Finally, 
we examined the relationship between the horizontal and vertical motions.
In Figure~\ref{fig:VB2Vbk},
the three horizontal speeds, 
$V_X$, $V_Y$, and $V_{XY}$, are plotted {\it vs.} 
$V_{\rm bk}^{\rm all}$ (left) and 
$V_{\rm bk}^{\rm inclined}$ (right)
in the upper three rows.
The last row shows 
$\langle|B|^{\rm all}\rangle$ {\it vs.} $V_{\rm bk}^{\rm all}$ (left) and
$\langle|B|^{\rm inclined}\rangle$ {\it vs.} $V_{\rm bk}^{\rm inclined}$ (right).
In all plots,
the individual edge-tracing results of the ME data are plotted in black triangles 
and those of the CEA data in red crosses.
The averages of the individual tracing results
are connected by thick lines to guide the eyes.
To distinguish the increasing and decreasing phases of
$\langle|B|^{\rm all}\rangle$ and $\langle|B|^{\rm inclined}\rangle$,
the thick lines corresponding to the two phases are plotted in different colors.
The growing and decaying phases of the ME results are plotted in black and blue,
respectively.
Those of the CEA results are plotted in red and orange,
respectively.
From the plots in the upper three rows,
we did not find a correlation or dependence between the horizontal motion and
the buoyant motion.
While the data points are not randomly scattered,
the horizontal and buoyant speeds seem to evolve independently of each other.
The last row, in contrast, shows a clear positive correlation between
$\langle|B|\rangle$ and $V_{\rm bk}$.
The slopes of the growing and decaying phases are slightly different.

\begin{figure}
\centerline{\includegraphics[width=1.0\textwidth,clip=]{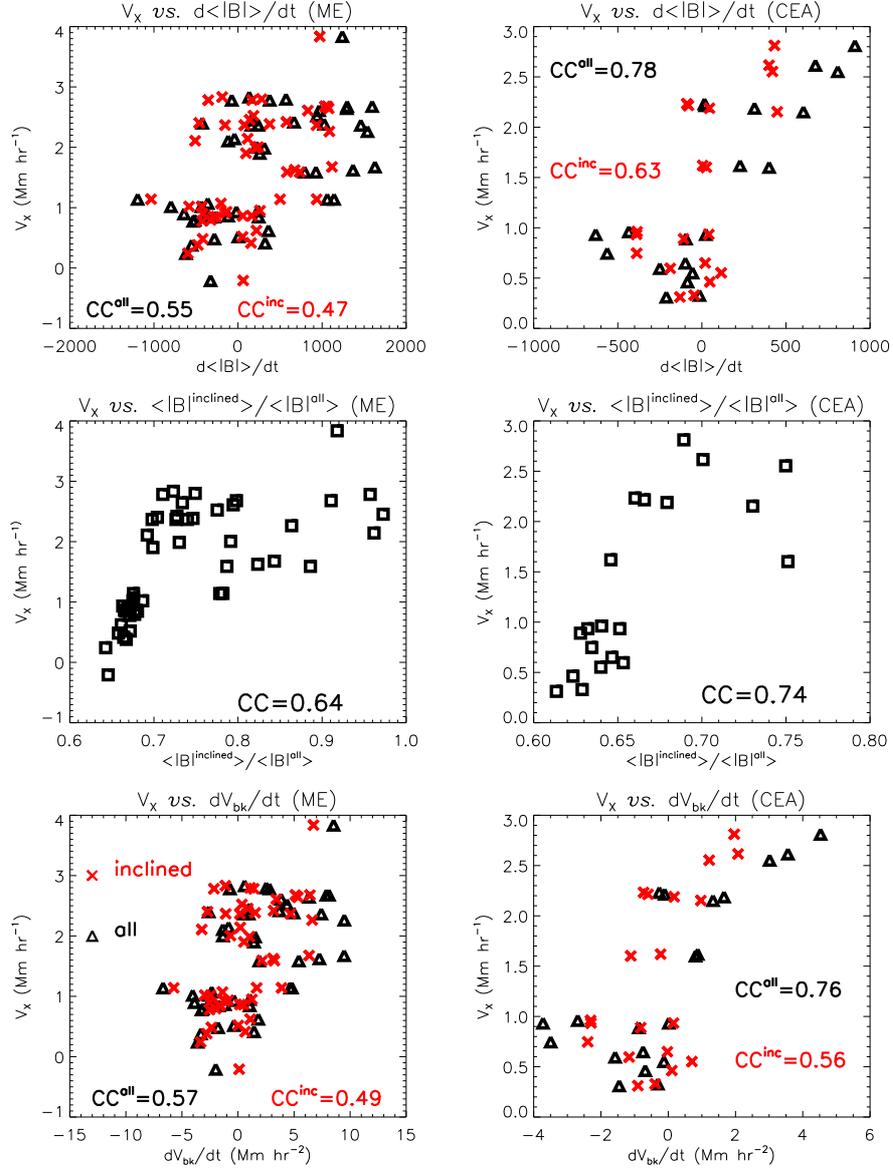}}
\caption{
$V_X$ {\it vs.} $d\langle|B|\rangle/dt$ (top row), 
$\langle|B|^{\rm inclined}\rangle/\langle|B|^{\rm all}\rangle/$ (middle row),
and $dV_{\rm bk}/dt$ (bottom row).
$d\langle|B|\rangle/dt$ 
is the notation for $d\langle|B|^{\rm all}\rangle/dt$ 
(black triangles)
and $d\langle|B|^{\rm inclined}\rangle/dt$ (red crosses),
and $dV_{\rm bk}/dt$ is the notation
for $dV_{\rm bk}^{\rm all}/dt$ (black triangles)
and $dV_{\rm bk}^{\rm inclined}/dt$ (red crosses).
The results of ME and CEA data are in the left and right columns,
respectively.
The computed correlation coefficients (CC) 
are shown in respective panels.
}
\label{fig:Vx2Brat}
\end{figure}

Since we found in Figures~\ref{fig:kin1} and \ref{fig:Bratio}
that
the temporal derivatives of 
the average field and buoyant velocities
and the ratio $\langle|B|^{\rm inclined}\rangle/\langle|B|^{\rm all}\rangle$
all decrease with time as $V_X$ does,
we investigated whether any correlation may exist among them.
In Figure~\ref{fig:Vx2Brat},
$V_X$ was plotted against $d\langle|B|\rangle/dt$ in the top row,
$\langle|B|^{\rm inclined}\rangle/\langle|B|^{\rm all}\rangle$ 
in the middle row,
and $dV_{\rm bk}/dt$ in the bottom row.
$d\langle|B|\rangle/dt$ is a notation to
represent $d\langle|B|^{\rm all}\rangle/dt$ and
$d\langle|B|^{\rm inclined}\rangle/dt$,
and $dV_{\rm bk}/dt$ represents 
$dV_{\rm bk}^{\rm all}/dt$ and $dV_{\rm bk}^{\rm inclined}/dt$.
The results from ME and CEA data are in the left and right columns,
respectively.
In the top and bottom rows, 
the inclined-field results are represented by red crosses,
and the all-direction results by black triangles.
Some positive linear correlations are visible in most of the plots,
especially in those in the right column.
To quantify the level of the correlation, 
we calculated and showed the correlation coefficients (CCs)
in the corresponding panels.
The results show that most of the correlation coefficients
are higher than 0.5,
indicating the existence of some positive correlation.
The correlation is stronger in the CEA data than in the ME data.
As described in Section~\ref{sec:data},
the CEA data do not include the earliest stage of flux emergence when
the flux is weak and can be contaminated by noise,
and the fields have been properly decomposed into radial and horizontal
components.
These facts can reduce the errors
in the determination of $\langle|B|^{\rm inclined}\rangle$,
$\langle|B|^{\rm all}\rangle$, $dX$, and $dY$,
and lead to the better correlation found in 
the CEA results.
The correlation coefficient also reveals that the correlation is stronger for
the all-direction quantities than for the inclined-field quantities.
The inclined fields were thought to be the top of emerging flux tube.
While the weaker magnitude and larger observational errors of 
the inclined field 
may partly contribute to the lower correlation,
it can also indicate that
the simplistic assumption of a single flux tube rising through the photosphere
may not be appropriate.

\section{Summary}
In this study,
we compared the observed horizontal motion and the
theoretically derived vertical motion of an emerging active region
to test
the assumption that it can be represented by 
the intersection of a rising magnetic tube,
whose dynamics is mainly governed by the magnetic buoyancy mechanism
and whose structure can be considered rigid
or not significantly distorted by surface and near-surface effects,
as it crosses the solar surface.

The selected target is AR~11645.
We tracked this AR emergence from 
the earliest detectable appearance of magnetic flux
at $\approx$~14:00~UT on 2 January 2013 until 23:59:59~UT
on 3 January 2013.
In this period of time,
the average field strength reached its peak at $\approx$ 
06:00~UT on 3 January 2013.
The region of emergence was initially free of preexisting field,
and was not associated with any eruption during the flux emergence process.
At the early stage, many small and transient flux concentrations rapidly emerged
with seemingly arbitrary polarity orientations,
and later organized in an East-West direction to form
a bipolar configuration.
\citet{chou1993ASPC} argued that this is an indication that
the EAR is not significantly affected by local surface effects.

Our analysis of the horizontal motion
showed that the separation of the two polarities of this region
was fastest ($V_X \approx 4.0$~Mm~hr$^{-1}$) at the beginning,
slowed down as the emergence continued,
and reached a near constant speed of $\le 1$~Mm~hr$^{-1}$ after
$\langle|B|^{\rm all}\rangle$ and 
$\langle|B|^{\rm inclined}\rangle$ reached their peaks.
The extension of the region
in the direction perpendicular to the line connecting the two poles
did not begin until $\approx$~4 hours after the first sign of flux emergence,
reached its peak velocity ($V_Y \approx 4.0$~Mm~hr$^{-1}$)
approximately three hours later,
and decreased to $\approx 1$~Mm hr$^{-1}$ at the end of the observation time.

To investigate the vertical motion,
we used the buoyant velocities of thin and thick flux-tube approximations
derived by \citet{CW1987SoPh},
and considered $dY/2$ as the flux-tube radius and 
the average $|B|$ of the EAR as the field strength
of the flux tube.
The computed buoyant velocities
revealed that the thin flux-tube approximation is inappropriate because
it results in an unreasonably high buoyant speed.
The thick flux-tube buoyant velocity of the inclined field vectors
has a similar magnitude as the
the horizontal velocity of this EAR, and is also consistent with
the Doppler velocity of inclined fields
reported by \citet{centeno2012ApJ}.
This indicates that the magnetic buoyancy mechanism is valid and
that the emergence was largely governed by it.

The temporal profiles of the average field strength and buoyant velocity
show a growing and a decaying phase,
and are positively correlated.
However, they do not show a dependence or correlation with
the observed horizontal motion.
The uncorrelation between the horizontal and vertical motions
suggests that the assumption that EARs are formed by
the emergence of flux tubes whose structures remain constant 
as they cross the surface should be taken with caution.

However, 
some positive correlations are found between
the separation velocity $V_X$ and
$d\langle|B|^{\rm all}\rangle/dt$, 
$dV_{\rm bk}^{\rm all}/dt$, and
$\langle|B|^{\rm inclined}\rangle/\langle|B|^{\rm all}\rangle$.
This indicates that the separation speed of the two polarities
can be related 
to the percentage of the inclined fields,
and the increase rate of the average field and
buoyant speed.
Whether or not this relationship is a general property for
emerging active regions would require more case studies.

%
\begin{acks}
This work is funded by the MOST of ROC under grant NSC 102-2112-M-008-018
and the MOE grant ``Aim for the Top University'' to
the National Central University.
\end{acks}

{\bf Disclosure of Potential Conflicts of Interest}

The authors declare that they have no conflicts of interest.

%
%
\bibliographystyle{spr-mp-sola}
\bibliography{ref}

\begin{thebibliography}{29}
\ifx\bisbn     \undefined \def\bisbn  #1{ISBN #1}\fi
\ifx\binits    \undefined \def\binits#1{#1}\fi
\ifx\bauthor   \undefined \def\bauthor#1{#1}\fi
\ifx\batitle   \undefined \def\batitle#1{#1}\fi
\ifx\bjtitle   \undefined \def\bjtitle#1{\textit{#1}}\fi
\ifx\bvolume   \undefined \def\bvolume#1{\textbf{#1}}\fi
\ifx\byear     \undefined \def\byear#1{#1}\fi
\ifx\bissue    \undefined \def\bissue#1{#1}\fi
\ifx\bfpage    \undefined \def\bfpage#1{#1}\fi
\ifx\blpage    \undefined \def\blpage #1{#1}\fi
\ifx\burl      \undefined \def\burl#1{\textsf{#1}}\fi
\ifx\href      \undefined \def\href#1#2{\textsf{#2}}\fi
\ifx\betal     \undefined \def\betal{\textit{et al.}}\fi
\ifx\bctitle   \undefined \def\bctitle#1{#1}\fi
\ifx\beditor   \undefined \def\beditor#1{#1}\fi
\ifx\bbtitle   \undefined \def\bbtitle#1{\textit{#1}}\fi
\ifx\bedition  \undefined \def\bedition#1{#1}\fi
\ifx\bseriesno \undefined \def\bseriesno#1{\textbf{#1}}\fi
\ifx\blocation \undefined \def\blocation#1{#1}\fi
\ifx\bsertitle \undefined \def\bsertitle#1{\textit{#1}}\fi
\ifx\bsnm      \undefined \def\bsnm#1{#1}\fi
\ifx\bsuffix   \undefined \def\bsuffix#1{#1}\fi
\ifx\bparticle \undefined \def\bparticle#1{#1}\fi
\ifx\barticle  \undefined \def\barticle#1{}\fi
\ifx\binstitute  \undefined \def\binstitute#1{#1}\fi
\ifx\bpublisher  \undefined \def\bpublisher#1{#1}\fi
\ifx\doiurl    \undefined
  \def\doiurl#1{\href{http://dx.doi.org/#1}{\textsf{DOI}}}\fi
\ifx\arxivurl  \undefined
  \def\arxivurl#1{\href{http://arxiv.org/abs/#1}{\textsf{arXiv}}}\fi
\ifx\adsurl    \undefined
  \def\adsurl#1{\href{http://adsabs.harvard.edu/abs/#1}{\textsf{ADS}}}\fi
\ifx\botherref \undefined \def\botherref#1{}\fi
\ifx\url       \undefined \def\url#1{\textsf{#1}}\fi
\ifx\bchapter  \undefined \def\bchapter#1{}\fi
\ifx\bbook     \undefined \def\bbook#1{}\fi
\ifx\bcomment  \undefined \def\bcomment#1{#1}\fi
\ifx\oauthor   \undefined \def\oauthor#1{#1}\fi
\ifx\citeauthoryear \undefined\def \citeauthoryear#1{#1}\fi
\ifx\endbibitem\undefined \def\endbibitem{}\fi
\ifx\bconflocation  \undefined \def\bconflocation#1{#1} \fi

\bibitem[\protect\citeauthoryear{{Brandenburg}
  \textit{et~al.}}{2011}]{Brandenburg2011ApJ}
\begin{barticle}
\bauthor{\bsnm{{Brandenburg}}, \binits{A.}},
\bauthor{\bsnm{{Kemel}}, \binits{K.}},
\bauthor{\bsnm{{Kleeorin}}, \binits{N.}},
\bauthor{\bsnm{{Mitra}}, \binits{D.}},
\bauthor{\bsnm{{Rogachevskii}}, \binits{I.}}:
\byear{2011},
\batitle{{Detection of Negative Effective Magnetic Pressure Instability in
  Turbulence Simulations}}.
\bjtitle{\apjl}
\bvolume{740},
\bfpage{L50}.
\doiurl{10.1088/2041-8205/740/2/L50}.
\adsurl{2011ApJ...740L..50B}.
\end{barticle}
\endbibitem

\bibitem[\protect\citeauthoryear{{Calabretta} and
  {Greisen}}{2002}]{CalabrettaGreisen2002AA}
\begin{barticle}
\bauthor{\bsnm{{Calabretta}}, \binits{M.R.}},
\bauthor{\bsnm{{Greisen}}, \binits{E.W.}}:
\byear{2002},
\batitle{{Representations of celestial coordinates in FITS}}.
\bjtitle{\aap}
\bvolume{395},
\bfpage{1077}.
\doiurl{10.1051/0004-6361:20021327}.
\adsurl{2002A\%26A...395.1077C}.
\end{barticle}
\endbibitem

\bibitem[\protect\citeauthoryear{{Caligari}, {Moreno-Insertis}, and
  {Sch\"ussler}}{1995}]{Caligari_etal1995ApJ}
\begin{barticle}
\bauthor{\bsnm{{Caligari}}, \binits{P.}},
\bauthor{\bsnm{{Moreno-Insertis}}, \binits{F.}},
\bauthor{\bsnm{{Sch\"ussler}}, \binits{M.}}:
\byear{1995},
\batitle{{Emerging flux tubes in the solar convection zone. 1: Asymmetry, tilt,
  and emergence latitude}}.
\bjtitle{\apj}
\bvolume{441},
\bfpage{886}.
\doiurl{10.1086/175410}.
\adsurl{1995ApJ...441..886C}.
\end{barticle}
\endbibitem

\bibitem[\protect\citeauthoryear{{Centeno}}{2012}]{centeno2012ApJ}
\begin{barticle}
\bauthor{\bsnm{{Centeno}}, \binits{R.}}:
\byear{2012},
\batitle{{The Naked Emergence of Solar Active Regions Observed with SDO/HMI}}.
\bjtitle{\apj}
\bvolume{759},
\bfpage{72}.
\doiurl{10.1088/0004-637X/759/1/72}.
\adsurl{2012ApJ...759...72C}.
\end{barticle}
\endbibitem

\bibitem[\protect\citeauthoryear{{Cheung} and {Isobe}}{2014}]{CI2014LRSP}
\begin{barticle}
\bauthor{\bsnm{{Cheung}}, \binits{M.C.M.}},
\bauthor{\bsnm{{Isobe}}, \binits{H.}}:
\byear{2014},
\batitle{{Flux Emergence (Theory)}}.
\bjtitle{Living Rev. in Solar Phys.}
\bvolume{11},
\bfpage{3}.
\doiurl{10.12942/lrsp-2014-3}.
\adsurl{2014LRSP...11....3C}.
\end{barticle}
\endbibitem

\bibitem[\protect\citeauthoryear{{Chintzoglou} and {Zhang}}{2013}]{CZ2013ApJ}
\begin{barticle}
\bauthor{\bsnm{{Chintzoglou}}, \binits{G.}},
\bauthor{\bsnm{{Zhang}}, \binits{J.}}:
\byear{2013},
\batitle{{Reconstructing the Subsurface Three-dimensional Magnetic Structure of
  a Solar Active Region Using SDO/HMI Observations}}.
\bjtitle{\apjl}
\bvolume{764},
\bfpage{L3}.
\doiurl{10.1088/2041-8205/764/1/L3}.
\adsurl{2013ApJ...764L...3C}.
\end{barticle}
\endbibitem

\bibitem[\protect\citeauthoryear{{Chou}}{1993}]{chou1993ASPC}
\begin{bchapter}
\bauthor{\bsnm{{Chou}}, \binits{D.-Y.}}:
\byear{1993},
\bctitle{{Structure of emerging flux regions.}}
In: \beditor{\bsnm{{Zirin}}, \binits{H.}},
\beditor{\bsnm{{Ai}}, \binits{G.}},
\beditor{\bsnm{{Wang}}, \binits{H.}} (eds.)
\bbtitle{IAU Colloq. 141: The Magnetic and Velocity Fields of Solar Active
  Regions},
\bsertitle{Astron. Soc. Pacific C. S.}
\bseriesno{46},
\bfpage{471}.
\adsurl{1993ASPC...46..471C}.
\end{bchapter}
\endbibitem

\bibitem[\protect\citeauthoryear{{Chou} and {Fisher}}{1989}]{CF1989ApJ}
\begin{barticle}
\bauthor{\bsnm{{Chou}}, \binits{D.-Y.}},
\bauthor{\bsnm{{Fisher}}, \binits{G.H.}}:
\byear{1989},
\batitle{{Dynamics of anchored flux tubes in the convection zone. I - Details
  of the model}}.
\bjtitle{\apj}
\bvolume{341},
\bfpage{533}.
\doiurl{10.1086/167514}.
\adsurl{1989ApJ...341..533C}.
\end{barticle}
\endbibitem

\bibitem[\protect\citeauthoryear{{Chou} and {Wang}}{1987}]{CW1987SoPh}
\begin{barticle}
\bauthor{\bsnm{{Chou}}, \binits{D.-Y.}},
\bauthor{\bsnm{{Wang}}, \binits{H.}}:
\byear{1987},
\batitle{{The separation velocity of emerging magnetic flux}}.
\bjtitle{\solphys}
\bvolume{110},
\bfpage{81}.
\doiurl{10.1007/BF00148204}.
\adsurl{1987SoPh..110...81C}.
\end{barticle}
\endbibitem

\bibitem[\protect\citeauthoryear{{Fan}}{2004}]{Fan2004ASPC}
\begin{bchapter}
\bauthor{\bsnm{{Fan}}, \binits{Y.}}:
\byear{2004},
\bctitle{{Dynamics of Emerging Flux Tubes}}.
In: \beditor{\bsnm{{Sakurai}}, \binits{T.}},
\beditor{\bsnm{{Sekii}}, \binits{T.}} (eds.)
\bbtitle{The Solar-B Mission and the Forefront of Solar Physics},
\bsertitle{Astron. Soc. Pacific C. S.}
\bseriesno{325},
\bfpage{47}.
\adsurl{2004ASPC..325...47F}.
\end{bchapter}
\endbibitem

\bibitem[\protect\citeauthoryear{{Fan}}{2008}]{Fan2008ApJ}
\begin{barticle}
\bauthor{\bsnm{{Fan}}, \binits{Y.}}:
\byear{2008},
\batitle{{The Three-dimensional Evolution of Buoyant Magnetic Flux Tubes in a
  Model Solar Convective Envelope}}.
\bjtitle{\apj}
\bvolume{676},
\bfpage{680}.
\doiurl{10.1086/527317}.
\adsurl{2008ApJ...676..680F}.
\end{barticle}
\endbibitem

\bibitem[\protect\citeauthoryear{{Harvey} and {Martin}}{1973}]{HM1973SoPh}
\begin{barticle}
\bauthor{\bsnm{{Harvey}}, \binits{K.L.}},
\bauthor{\bsnm{{Martin}}, \binits{S.F.}}:
\byear{1973},
\batitle{{Ephemeral Active Regions}}.
\bjtitle{\solphys}
\bvolume{32},
\bfpage{389}.
\doiurl{10.1007/BF00154951}.
\adsurl{1973SoPh...32..389H}.
\end{barticle}
\endbibitem

\bibitem[\protect\citeauthoryear{{Hoeksema}
  \textit{et~al.}}{2014}]{HMI_ME2014SoPh289}
\begin{barticle}
\bauthor{\bsnm{{Hoeksema}}, \binits{J.T.}},
\bauthor{\bsnm{{Liu}}, \binits{Y.}},
\bauthor{\bsnm{{Hayashi}}, \binits{K.}},
\bauthor{\bsnm{{Sun}}, \binits{X.}},
\bauthor{\bsnm{{Schou}}, \binits{J.}},
\bauthor{\bsnm{{Couvidat}}, \binits{S.}},
\bauthor{\bsnm{{Norton}}, \binits{A.}},
\bauthor{\bsnm{{Bobra}}, \binits{M.}},
\bauthor{\bsnm{{Centeno}}, \binits{R.}},
\bauthor{\bsnm{{Leka}}, \binits{K.D.}},
\bauthor{\bsnm{{Barnes}}, \binits{G.}},
\bauthor{\bsnm{{Turmon}}, \binits{M.}}:
\byear{2014},
\batitle{{The Helioseismic and Magnetic Imager (HMI) Vector Magnetic Field
  Pipeline: Overview and Performance}}.
\bjtitle{\solphys}
\bvolume{289},
\bfpage{3483}.
\doiurl{10.1007/s11207-014-0516-8}.
\adsurl{2014SoPh..289.3483H}.
\end{barticle}
\endbibitem

\bibitem[\protect\citeauthoryear{{Kubo}, {Shimizu}, and
  {Lites}}{2003}]{ksl2003ApJ}
\begin{barticle}
\bauthor{\bsnm{{Kubo}}, \binits{M.}},
\bauthor{\bsnm{{Shimizu}}, \binits{T.}},
\bauthor{\bsnm{{Lites}}, \binits{B.W.}}:
\byear{2003},
\batitle{{The Evolution of Vector Magnetic Fields in an Emerging Flux Region}}.
\bjtitle{\apj}
\bvolume{595},
\bfpage{465}.
\doiurl{10.1086/377333}.
\adsurl{2003ApJ...595..465K}.
\end{barticle}
\endbibitem

\bibitem[\protect\citeauthoryear{{Leka}
  \textit{et~al.}}{1996}]{Leka_etal1996ApJ}
\begin{barticle}
\bauthor{\bsnm{{Leka}}, \binits{K.D.}},
\bauthor{\bsnm{{Canfield}}, \binits{R.C.}},
\bauthor{\bsnm{{McClymont}}, \binits{A.N.}},
\bauthor{\bsnm{{van Driel-Gesztelyi}}, \binits{L.}}:
\byear{1996},
\batitle{{Evidence for Current-carrying Emerging Flux}}.
\bjtitle{\apj}
\bvolume{462},
\bfpage{547}.
\doiurl{10.1086/177171}.
\adsurl{1996ApJ...462..547L}.
\end{barticle}
\endbibitem

\bibitem[\protect\citeauthoryear{{Lites}, {Skumanich}, and {Martinez
  Pillet}}{1998}]{lsmp1998AA}
\begin{barticle}
\bauthor{\bsnm{{Lites}}, \binits{B.W.}},
\bauthor{\bsnm{{Skumanich}}, \binits{A.}},
\bauthor{\bsnm{{Martinez Pillet}}, \binits{V.}}:
\byear{1998},
\batitle{{Vector magnetic fields of emerging solar flux. I. Properties at the
  site of emergence}}.
\bjtitle{\aap}
\bvolume{333},
\bfpage{1053}.
\adsurl{1998A\%26A...333.1053L}.
\end{barticle}
\endbibitem

\bibitem[\protect\citeauthoryear{{Metcalf}}{1994}]{Metcalf1994SoPh}
\begin{barticle}
\bauthor{\bsnm{{Metcalf}}, \binits{T.R.}}:
\byear{1994},
\batitle{{Resolving the 180-degree ambiguity in vector magnetic field
  measurements: The 'minimum' energy solution}}.
\bjtitle{\solphys}
\bvolume{155},
\bfpage{235}.
\doiurl{10.1007/BF00680593}.
\adsurl{1994SoPh..155..235M}.
\end{barticle}
\endbibitem

\bibitem[\protect\citeauthoryear{{Parker}}{1955}]{parker1955ApJ121}
\begin{barticle}
\bauthor{\bsnm{{Parker}}, \binits{E.N.}}:
\byear{1955},
\batitle{{The Formation of Sunspots from the Solar Toroidal Field.}}
\bjtitle{\apj}
\bvolume{121},
\bfpage{491}.
\doiurl{10.1086/146010}.
\adsurl{1955ApJ...121..491P}.
\end{barticle}
\endbibitem

\bibitem[\protect\citeauthoryear{{Parker}}{1978}]{parker1978ApJ}
\begin{barticle}
\bauthor{\bsnm{{Parker}}, \binits{E.N.}}:
\byear{1978},
\batitle{{Hydraulic concentration of magnetic fields in the solar photosphere.
  VI - Adiabatic cooling and concentration in downdrafts}}.
\bjtitle{\apj}
\bvolume{221},
\bfpage{368}.
\doiurl{10.1086/156035}.
\adsurl{1978ApJ...221..368P}.
\end{barticle}
\endbibitem

\bibitem[\protect\citeauthoryear{{Pesnell}, {Thompson}, and
  {Chamberlin}}{2012}]{SDO_2012SoPh275}
\begin{barticle}
\bauthor{\bsnm{{Pesnell}}, \binits{W.D.}},
\bauthor{\bsnm{{Thompson}}, \binits{B.J.}},
\bauthor{\bsnm{{Chamberlin}}, \binits{P.C.}}:
\byear{2012},
\batitle{{The Solar Dynamics Observatory (SDO)}}.
\bjtitle{\solphys}
\bvolume{275},
\bfpage{3}.
\doiurl{10.1007/s11207-011-9841-3}.
\adsurl{2012SoPh..275....3P}.
\end{barticle}
\endbibitem

\bibitem[\protect\citeauthoryear{{Rempel}}{2011}]{Rempel2011ApJ}
\begin{barticle}
\bauthor{\bsnm{{Rempel}}, \binits{M.}}:
\byear{2011},
\batitle{{Subsurface Magnetic Field and Flow Structure of Simulated Sunspots}}.
\bjtitle{\apj}
\bvolume{740},
\bfpage{15}.
\doiurl{10.1088/0004-637X/740/1/15}.
\adsurl{2011ApJ...740...15R}.
\end{barticle}
\endbibitem

\bibitem[\protect\citeauthoryear{{Rempel} and
  {Cheung}}{2014}]{RempelCheung2014ApJ}
\begin{barticle}
\bauthor{\bsnm{{Rempel}}, \binits{M.}},
\bauthor{\bsnm{{Cheung}}, \binits{M.C.M.}}:
\byear{2014},
\batitle{{Numerical Simulations of Active Region Scale Flux Emergence: From
  Spot Formation to Decay}}.
\bjtitle{\apj}
\bvolume{785},
\bfpage{90}.
\doiurl{10.1088/0004-637X/785/2/90}.
\adsurl{2014ApJ...785...90R}.
\end{barticle}
\endbibitem

\bibitem[\protect\citeauthoryear{{Schou}
  \textit{et~al.}}{2012}]{HMI_2012SoPh275}
\begin{barticle}
\bauthor{\bsnm{{Schou}}, \binits{J.}},
\bauthor{\bsnm{{Scherrer}}, \binits{P.H.}},
\bauthor{\bsnm{{Bush}}, \binits{R.I.}},
\bauthor{\bsnm{{Wachter}}, \binits{R.}},
\bauthor{\bsnm{{Couvidat}}, \binits{S.}},
\bauthor{\bsnm{{Rabello-Soares}}, \binits{M.C.}},
\bauthor{\bsnm{{Bogart}}, \binits{R.S.}},
\bauthor{\bsnm{{Hoeksema}}, \binits{J.T.}},
\bauthor{\bsnm{{Liu}}, \binits{Y.}},
\bauthor{\bsnm{{Duvall}}, \binits{T.L.}},
\bauthor{\bsnm{{Akin}}, \binits{D.J.}},
\bauthor{\bsnm{{Allard}}, \binits{B.A.}},
\bauthor{\bsnm{{Miles}}, \binits{J.W.}},
\bauthor{\bsnm{{Rairden}}, \binits{R.}},
\bauthor{\bsnm{{Shine}}, \binits{R.A.}},
\bauthor{\bsnm{{Tarbell}}, \binits{T.D.}},
\bauthor{\bsnm{{Title}}, \binits{A.M.}},
\bauthor{\bsnm{{Wolfson}}, \binits{C.J.}},
\bauthor{\bsnm{{Elmore}}, \binits{D.F.}},
\bauthor{\bsnm{{Norton}}, \binits{A.A.}},
\bauthor{\bsnm{{Tomczyk}}, \binits{S.}}:
\byear{2012},
\batitle{{Design and Ground Calibration of the Helioseismic and Magnetic Imager
  (HMI) Instrument on the Solar Dynamics Observatory (SDO)}}.
\bjtitle{\solphys}
\bvolume{275},
\bfpage{229}.
\doiurl{10.1007/s11207-011-9842-2}.
\adsurl{2012SoPh..275..229S}.
\end{barticle}
\endbibitem

\bibitem[\protect\citeauthoryear{{Sch\"ussler}}{1979}]{Schuessler1979AA}
\begin{barticle}
\bauthor{\bsnm{{Sch\"ussler}}, \binits{M.}}:
\byear{1979},
\batitle{{Magnetic buoyancy revisited - Analytical and numerical results for
  rising flux tubes}}.
\bjtitle{\aap}
\bvolume{71},
\bfpage{79}.
\adsurl{1979A\%26A....71...79S}.
\end{barticle}
\endbibitem

\bibitem[\protect\citeauthoryear{{Shimizu}
  \textit{et~al.}}{2002}]{shimizu_etal2002ApJ}
\begin{barticle}
\bauthor{\bsnm{{Shimizu}}, \binits{T.}},
\bauthor{\bsnm{{Shine}}, \binits{R.A.}},
\bauthor{\bsnm{{Title}}, \binits{A.M.}},
\bauthor{\bsnm{{Tarbell}}, \binits{T.D.}},
\bauthor{\bsnm{{Frank}}, \binits{Z.}}:
\byear{2002},
\batitle{{Photospheric Magnetic Activities Responsible for Soft X-Ray Pointlike
  Microflares. I. Identifications of Associated Photospheric/Chromospheric
  Activities}}.
\bjtitle{\apj}
\bvolume{574},
\bfpage{1074}.
\doiurl{10.1086/340998}.
\adsurl{2002ApJ...574.1074S}.
\end{barticle}
\endbibitem

\bibitem[\protect\citeauthoryear{{Strous}
  \textit{et~al.}}{1996}]{strous_etal1996AA}
\begin{barticle}
\bauthor{\bsnm{{Strous}}, \binits{L.H.}},
\bauthor{\bsnm{{Scharmer}}, \binits{G.}},
\bauthor{\bsnm{{Tarbell}}, \binits{T.D.}},
\bauthor{\bsnm{{Title}}, \binits{A.M.}},
\bauthor{\bsnm{{Zwaan}}, \binits{C.}}:
\byear{1996},
\batitle{{Phenomena in an emerging active region. I. Horizontal dynamics.}}
\bjtitle{\aap}
\bvolume{306},
\bfpage{947}.
\adsurl{1996A\%26A...306..947S}.
\end{barticle}
\endbibitem

\bibitem[\protect\citeauthoryear{{Sun}}{2013}]{Sun2013arXiv}
\begin{botherref}
\oauthor{\bsnm{{Sun}}, \binits{X.}}:
2013,
{On the Coordinate System of Space-Weather HMI Active Region Patches (SHARPs):
  A Technical Note}.
\textit{ArXiv e-prints}.
\adsurl{2013arXiv1309.2392S}.
\end{botherref}
\endbibitem

\bibitem[\protect\citeauthoryear{{Tanaka}}{1991}]{Tanaka1991SoPh}
\begin{barticle}
\bauthor{\bsnm{{Tanaka}}, \binits{K.}}:
\byear{1991},
\batitle{{Studies on a very flare-active delta group - Peculiar delta spot
  evolution and inferred subsurface magnetic rope structure}}.
\bjtitle{\solphys}
\bvolume{136},
\bfpage{133}.
\doiurl{10.1007/BF00151700}.
\adsurl{1991SoPh..136..133T}.
\end{barticle}
\endbibitem

\bibitem[\protect\citeauthoryear{{Weber}, {Fan}, and
  {Miesch}}{2011}]{Weber_etal2011ApJ}
\begin{barticle}
\bauthor{\bsnm{{Weber}}, \binits{M.A.}},
\bauthor{\bsnm{{Fan}}, \binits{Y.}},
\bauthor{\bsnm{{Miesch}}, \binits{M.S.}}:
\byear{2011},
\batitle{{The Rise of Active Region Flux Tubes in the Turbulent Solar
  Convective Envelope}}.
\bjtitle{\apj}
\bvolume{741},
\bfpage{11}.
\doiurl{10.1088/0004-637X/741/1/11}.
\adsurl{2011ApJ...741...11W}.
\end{barticle}
\endbibitem

\end{thebibliography}

%

%
%
%
%

\end{article} 
\end{document}